\documentclass{aa}  
\usepackage{amsmath}
\usepackage{soul} % para tachar palabras
\usepackage{xcolor}

\usepackage{graphicx}
\usepackage{txfonts}
\begin{document}

  \title{Planetesimal fragmentation and giant planet formation II:}

   \subtitle{dependencies with planetesimal relative velocities and compositions}

   \author{I. L. San Sebasti\'an \inst{1,2}\thanks{\email{irina@fcaglp.unlp.edu.ar}}, O. M. Guilera \inst{1,3,4} \and M. G. Parisi \inst{1,5}
   }

   \institute{Facultad de Ciencias Astron\'omicas y Geof\'{\i}sicas, UNLP, Paseo del Bosque S/N, B1900FWA La Plata, Argentina.
     \and CONICET, Godoy Cruz 2290, CABA, Argentina. 
     \and Instituto de Astrof\'{\i}sica de La Plata, CCT La Plata, CONICET-UNLP, Paseo del Bosque S/N, B1900FWA La Plata, Argentina.
     \and Instituto de Astrof\'{\i}sica, Pontificia Universidad Catolica de Chile, Santiago, Chile.
     \and Instituto Argentino de Radioastronom\'{\i}a, CCT-La Plata, CONICET-CICPBA, CC N. 5 (1894), Villa Elisa, Argentina.
   }

   \date{Received xxx; accepted xxx}

% \abstract{}{}{}{}{} 
% 5 {} token are mandatory

 \abstract
  % context heading (optional)
  % {} leave it empty if necessary  
   {Most of planet formation models that incorporate planetesimal fragmentation consider a catastrophic impact energy threshold for basalts at a constant velocity of 3~km/s during all the process of the formation of the planets. However, as planets grow the relative velocities of the surrounding planetesimals increase from velocities of the order of m/s to a few km/s. In addition, beyond the ice line where giant planets are formed, planetesimals are expected to be composed  roughly by 50 \% of ices.}
  % aims heading (mandatory)
   {We aim to study the role of planetesimal fragmentation on giant planet formation considering planetesimal catastrophic impact energy threshold as a function of the planetesimal relative velocities and compositions.}
  % methods heading (mandatory)
   {We improve our model of planetesimal fragmentation incorporating a functional form of the catastrophic impact energy threshold with the planetesimal relative velocities and compositions. We also improve in our model the accretion of small fragments produced by the fragmentation of planetesimals during the collisional cascade considering specific pebble accretion rates.}
  % results heading (mandatory)
   {We find that a more accurate and realistic model for the calculation of the catastrophic impact energy threshold tends to slow down the formation of massive cores. Only for reduced grain opacity values at the envelope of the planet, the cross-over mass is achieved before the disk time-scale dissipation.}
  % conclusions heading (optional), leave it empty if necessary 
   {While planetesimal fragmentation favors the quick formation of massive cores of $5-10~\text{M}_{\oplus}$ the cross-over mass could be inhibited by planetesimal fragmentation. However, grain opacity reduction or pollution by the accreted planetesimals together with planetesimal fragmentation could explain the formation of giant planets with low-mass cores.} 
 

   \maketitle
%
%-------------------------------------------------------------------

\section{Introduction}
\label{sec1}

The standard model of giant planet formation is the core-accretion mechanism \citep{Mizuno1980, Stevenson1982, Pollack1996, Guilera2010, Helled2014}. According to this model, the planet forms from accretion of planetesimals onto a solid core until it has enough mass to start accreting gas from the protoplanetary disk. When the cross-over mass is reached, the mass of the envelope equals the core mass (at $\sim 10 M_{\oplus}$), and the planet starts to accrete big quantities of gas in a short period of time, this process is known as runaway gas  accretion. At some point, for mechanisms not yet well understood, the accretion of gas onto the planet is limited. Finally, the planet evolves cooling and contracting at constant mass.

There is an alternative model developed in the past few years, also based in the accretion of solid material, for the formation of giant planets. This model is based on the accretion of small sized particles (called pebbles) for the formation of the planet core. Unlike planetesimals, pebbles can be accreted by the full Hill sphere making their accretion rates significantly larger than planetesimal accretion rates \citep{OrmelKlahr2010, LJ2012}. 

In the standard core accretion mechanism an important issue is the formation of a massive core before the dissipation of the gaseous component of the protoplanetary disk. In this framework, the process of accretion of planetesimals plays a fundamental role. As the planetary embryo grows, the surrounding planetesimals increase their relative velocities due to the gravitational stirring produced by the embryo, leading to disruptive collisions among them. This effect produces a cascade of collision fragments. \cite{Inaba2003}, \cite{Kobayashi2011}, \cite{OrmelKobayashi2012}, and \citet{Guilera2014} found that excessive fragmentation may cause that the smaller fragments produced by these collisions drift inwards by gas drag and a significant amount of mass could be lost stalling the oligarchic growth. Moreover, when the core reaches several Earth masses the relative planetesimal velocities become high enough to produce supercatastrophic collisions among planetesimals \citep{Guilera2014,Chambers2014}. In this case, \citet{Guilera2014} considered that the planetesimals involved in such collision are pulverized and the mass of such bodies is lost. On the other hand, \cite{Chambers2014} found that small planetesimals collide frequently and produce rapid embryo growth.  Unlike \cite{Guilera2014}, \cite{Chambers2014} considered that the mass that would go into fragments smaller than the minimum size considered in the model (referred as pebbles of second generation) is assigned to become an equivalent mass of the minimum size adopted in their model, since they assume that tiny fragments quickly coagulate into new pebbles.

Despite the different planetesimal fragmentation models, during the process of planetary formation, fragmentation of planetesimals is an important effect that may inhibit or favor the formation of giant planet cores. In a collision between two planetesimals, the catastrophic impact energy threshold required to fragment and disperse fifty percent of the target mass (also known as specific impact energy) is an important function that has to be defined in models of planetesimal fragmentation. This function depends on many factors of the collision, such as the impact velocity, the planetesimal type of material and its porosity, the sizes of the planetesimals and other factors that may affect the outcome of the collision \citep[see][]{Jutzi2015}. For example, \citet{BenzAsphaug1999} found that a target of a fixed size compose by ice-material has catastrophic impact energy threshold lower than if it were composed by basalt. \cite{Benz2000} also found that low velocity collisions are more efficient in destroying and dispersing basaltic bodies than collisions at higher velocities. \cite{Jutzi2010} studied from SPH simulations the porosity of the planetesimals in a collision and found that, in the strength regime, porous targets are more difficult to disrupt that non-porous ones while in the gravity regime, the outcome of the collision for porous targets depends on gravity and porosity, unlike non-porous ones that are also controlled by the strength. Recently, \citet{Beitz-et-al-2011} and \citet{Bukhari-et-al-2017} showed in low-velocity impact experiments that the strength of compacted dust aggregates is much weaker than that of porous rocks. This outcome is analogous to \citet{Jutzi2010}. It is important to remark here, that most models of planet formation that incorporate the planetesimal fragmentation process use the prescription of \cite{BenzAsphaug1999} for the catastrophic impact energy threshold for non-porous basalts at impact velocities of $3$~km/s \citep{Kobayashi2011, OrmelKobayashi2012, Guilera2014, Chambers2014}. However, \citet{Guilera2014} showed that during the formation of massive cores the relative planetesimal velocities are incremented from cm/sec--m/sec to km/sec as planet grows.  

In our previous work \citep{Guilera2014}, we incorporated a model of planetesimal fragmentation into our global model of giant planet formation \citep{Guilera2010}. We found that the formation of massive cores are only possible from an initial population of big planetesimals and massive disks. In other situations, planetesimal fragmentation tends to inhibit the formation of massive cores before the dissipation of the disk. In this new work, we incorporate to our planetesimal fragmentation model the dependence of the catastrophic impact energy threshold with material composition and planetesimal relative velocities. We also incorporate different velocity regimes for planetesimal relative velocities (Keplerian shear and dispersion dominate regimes) in the planetesimal fragmentation model, and pebbles accretion rates for the small particles product of the collision process into the global model of giant planet formation. The aim of this work is to analyze the impact onto the formation of massive cores of the new phenomena incorporated, specially planetesimals relative velocities and planetesimal compositions. 

This paper is organized as follows: in Sec.~\ref{sec2}, we describe the improvements of our model of giant planet formation and the improvements of the model of planetesimal fragmentation; in Sec.~\ref{sec3}, we present the results of our simulations for the formation of a giant planet, and finally, in Sec.\ref{sec4} we present the summary and conclusions of our results.

%--------------------------------------------------------------------
\section{Our model of giant planet formation}
\label{sec2}

In a series of previous works \citep{Guilera2010, Guilera2014} we developed a numerical model that describes the formation of giant planets immersed in a protoplanetary disk that evolves in time. In our model, the  protoplanetary disk is represented by a gaseous and a solid components. Planets grow by the simultaneous accretion of solids and gas. The solid component of the disk evolves by planet accretion, radial drift due to nebular drag and collisional evolution, while the gaseous component evolves by an exponential decay (see Appendix \ref{sec5-1} for technical details on the model).

 In this work we improve our planetesimal fragmentation model developed in \citet{Guilera2014} incorporating a dependence of $Q^*_d$, the catastrophic impact energy threshold, with the planetesimal relative velocities and the planetesimal composition considering a mix of non-porous ices and, as a proxy for the rocky material of planetesimals, non-porous basalts. We also incorporate pebble accretion rates for the small fragments product of the planetesimal fragmentation process (hereafter, pebbles of second generation), and different velocity regime models for the calculation of low and high relative planetesimal velocities. 

 The improvements in our model are presented in the next subsections.

\subsection{Improvements in the solid accretion rates}
\label{sec2-1}

In the past few years, a new model of solid particles accretion has been proposed as an alternative of the core accretion model for the formation of solid massive cores. Such model proposes that massive cores, precursors of giant planet cores, can be quickly formed by 100-1000 kilometers embryos accreting cm-sized particles known as pebbles. \citet{OrmelKlahr2010} and \citet{LJ2012} found that pebbles, particles with Stokes numbers below the unity ($S_{t} \lesssim1$), are strongly coupled to the gas and can be very efficiently accreted by hundred-thousand sized embryos. While planetesimals can be accreted by a fraction of the Hill radius of the growing planet, pebbles can be accreted by the full Hill radius, making pebble accretion rates significantly larger than planetesimal accretion rates. 

 However, the dominant initial size of the accreted solids (cm-sized particles named pebbles or 1-100 km-sized bodies named planetesimals) is unknown \citep{Helled2014,Johansen2014}.  At the begging of our simulations, the solid component of the protoplanetary disk is composed by planetesimals of radius 100 km with the distribution given by the framework of the oligarchic growth \citep{KokuboIda1998,KokuboIda2000,KokuboIda2002,IdaMakino1993}. Thus, we assume that there are no initial pebbles in our model. However, pebbles appear as result of the planetesimal collisional evolution, i.e., second generation pebbles.

Following \citet{Guilera2016} and \citet{Guilera-Sandor2017}, we improve the solid accretion rates by incorporating the pebble accretion rate for pebbles of second generation in addition to the planetesimals accretion.  For planetesimals we adopt the accretion rates given by  \cite{Inaba2001} and for pebbles of second generation we use the pebble accretion rates given by \cite{Lambrechts2014} with a reduction factor that takes into account the scale height of the pebbles in comparison with the Hill radius of the planet (see Appendix \ref{sec5-2}).

\subsection{Improvements to our planetesimal fragmentation model}
\label{sec2-2}

Our planetesimal fragmentation model, which is based in the Boulder Code (\cite{Morbidelli2009} and Supplementary Material), describes the collisional evolution of the planetesimal population. A brief description of the planetesimal fragmentation model and technical details are presented in Appendix \ref{sec6}).

\subsubsection{Velocities and probabilities of collision regimes}
\label{sec-2-2-1}

Following \citet{Morbidelli2009}, the impact rate between targets and projectiles estimated in our model is given by  
\begin{eqnarray}
  \text{Impact rate}|_{\text{A}}= \frac {\alpha V_{rel}}{4 H a (\delta a + 2 a e_p)} E \left(1 + b \frac{V_{esc}^2}{V_{rel}^2}\right) (R_T + R_P)^2,  
  \label{eq1-sec-2-2-1}
\end{eqnarray}
where $V_{rel}$ is the relative velocity between targets and projectiles, $\alpha$ is a coefficient that depends on $V_{rel}$, $H$ is the mutual scale height, $a$ and $\delta a$ are the mean semi-major axis and the width of the annulus that contains the targets and projectiles, respectively, $e_p$ is the eccentricity of the projectiles, $E$ is a coefficient that considers the deviation of the gravitational focusing corresponding to the two-body problem at low relative velocities, $b$ is also a function of $V_{rel}$, and $V_{esc}$ is the mutual escape velocity. Finally, $R_T$ and $R_P$ are the radius of the targets and projectiles, respectively \citep[see][for details]{Morbidelli2009}.  

It is important to note, that the relative velocity among planetesimals used in the Boulder code is the dispersion velocity (the velocity corresponding to the dispersion regime). However, when the planetesimal relative velocity tends to zero, the impact rate between them is undetermined. \cite{Weidenschilling2011} noticed this problem and incorporated the relative velocity corresponding to the keplerian shear. Following \cite{Weidenschilling2011}, we incorporate in our planetesimal fragmentation model different velocity regimes and probabilities of collision for lower relative velocities given by \cite{Greenberg1991} to calculate collision rates more accurately. We consider three different regimes and their transitions according to \cite{Greenberg1991}, regime A: dominance by random motion (impact rate given by Eq.\ref{eq1-sec-2-2-1}); regime B: dominance by Keplerian shear motion; regime C: Keplerian shear dominance in a very thin disk (see Appendix \ref{sec6-2} for details).

\subsubsection{Catastrophic impact energy threshold}
\label{sec2-2-2}

\begin{figure}
  \centering
  \includegraphics[width=\hsize]{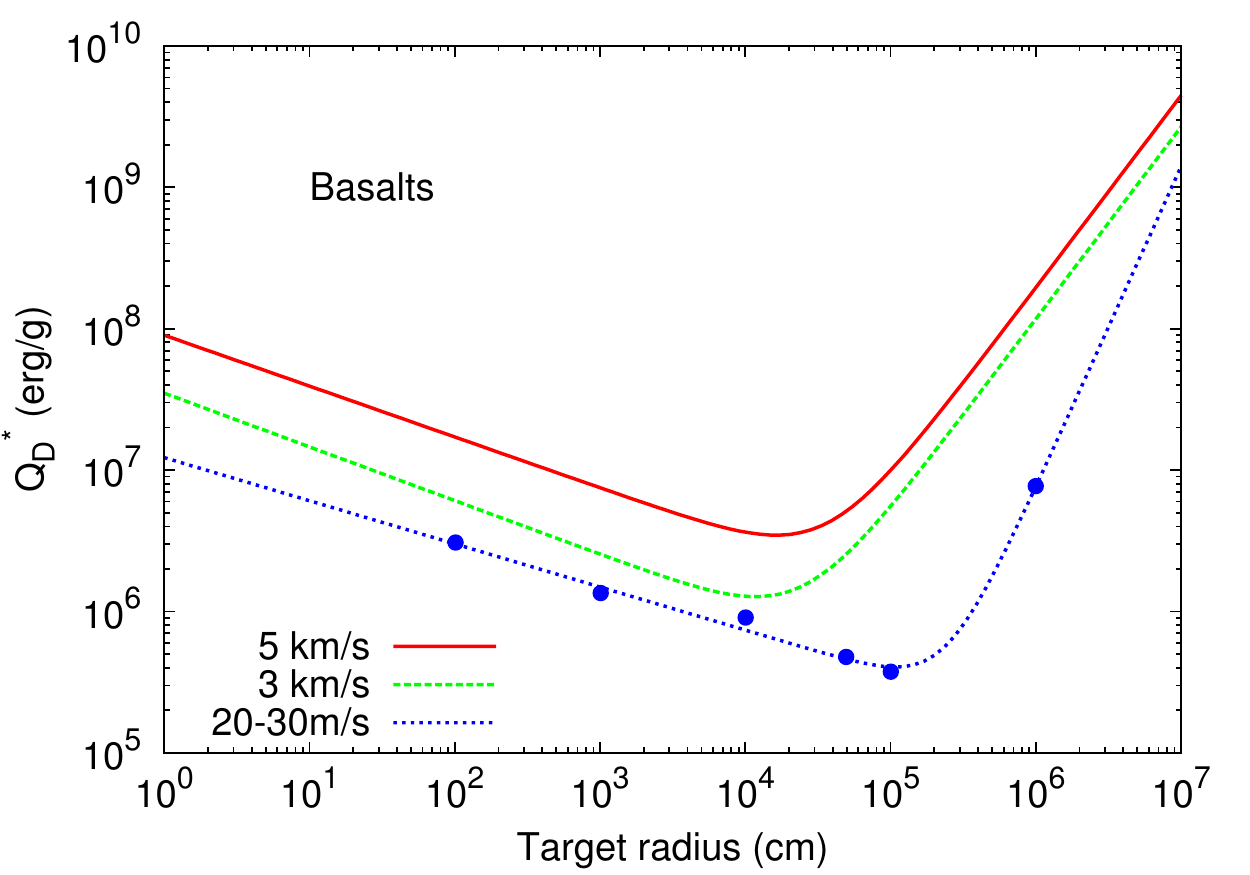}
  \caption{Catastrophic disruption thresholds for basalt targets at impact velocities of 20-30 m/s \citep{Benz2000}, 3km/s and 5km/s \citep{BenzAsphaug1999}. The blue points correspond to the discrete data extracted from \citep{Benz2000}.
  }
  \label{basalts}
\end{figure}
\begin{figure}
  \centering
  \includegraphics[width=\hsize]{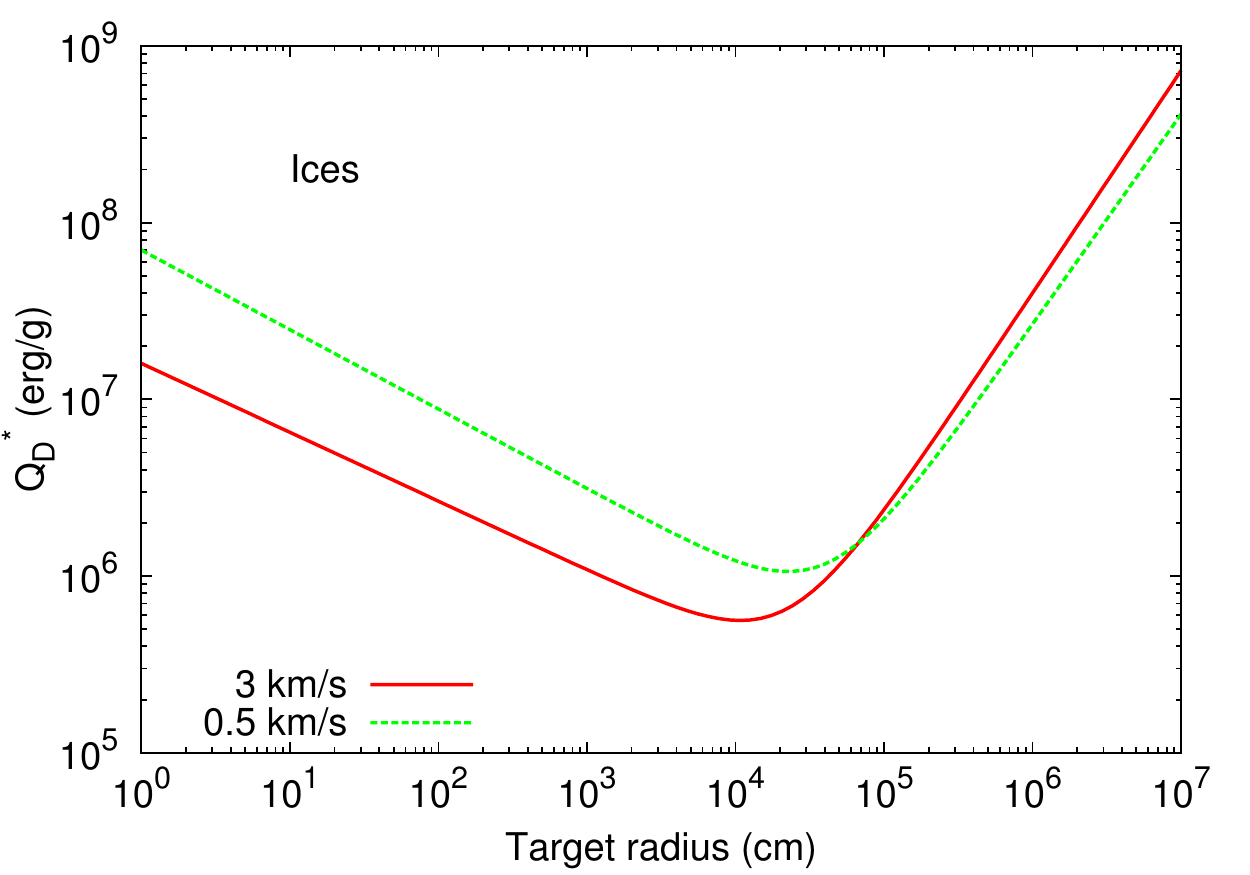}
  \caption{Catastrophic disruption thresholds for icy targets at impact velocities of 0.5 km/s and 3 km/s \citep{BenzAsphaug1999}. 
  }
  \label{ice}
\end{figure}

The catastrophic impact energy threshold per unit target mass $Q_{D}^*$ is the energy needed to fragment and disperse half of the target mass in an impact, i.e., the threshold for catastrophic disruption. This quantity plays an important role in the collisional evolution of the planetesimal population. As we mentioned in Sec.~\ref{sec1}, $Q_{D}^*$ depends on many factors of the collision; particularly important are the relative velocity among planetesimals, which determines the impact velocity of the collision, and the planetesimals composition.

Fig.~\ref{basalts} represents $Q_{D}^*$ as a function of the radius of a non-porous basalt target for three different values of the impact velocity ($\sim 25$ m/s, 3 km/s and 5km/s), while Fig.\ref{ice} represents this threshold for a non-porous icy target for two values of the impact velocity (0.5 km/s and 3 km/s) (see Appendix \ref{sec6-3}). For basalt targets, we can see that for a fixed target's radius, the smaller the impact velocity, the smaller $Q_{D}^*$. This phenomenon can have an important effect during the formation of massive cores due to the fact that, initially, planetesimal relative velocities are low and are then incremented as the planet grows. We remark again that in most works of planetary formation that include planetesimal fragmentation it is used $Q_{d}^*$ for non-porous basalt targets at an impact velocity of 3 km/s \citep{OrmelKobayashi2012, Chambers2014}. Given the functional dependency of $Q_{d}^*$ with target's radius for different impact velocities given by \cite{BenzAsphaug1999} and \cite{Benz2000}, for a fixed target's radius, we implemented an interpolation between the impact velocity dependencies to get an improved value of $Q_{d}^*$ as a function of the impact velocity. We also include an interpolation using the different curves of $Q^*_d$ of \cite{BenzAsphaug1999} for different velocities for ices. We note that if the impact velocities are greater or lower than the velocities corresponding to the upper and lower curves of $Q_{d}^*$ adopted, we do not extrapolate $Q_{d}^*$. In that case, we adopt $Q_{d}^*$ corresponding to the maximum or minimum velocity used in \cite{BenzAsphaug1999} and \cite{Benz2000}.

On the other hand, according to \cite{Lodders2003}, beyond the iceline the amount of solid mass in the protoplanetary disk increases by a factor of two meaning that fifty percent of the material behind the iceline should be condensated in the primitive Solar System. The ice-to-rock ratio derived from trans-Neptunian objects, comets and irregular satellites of giant planets \citep{McDonnell1987,Stern1997,JohnsonLunine2005}  confirms this.
 As we are interested in the formation of giant planets behind the iceline, we assume that planetesimals are composed by ices and basalts following \cite{Lodders2003}. Finally, to implement the dependence of $Q^*_D$ with the impact velocity and the target's composition, we first interpolate between the curves of $Q_{d}^*$ as a function of the target's radius for the different impact velocities, obtaining the values of $Q^*_D$ for each pure material (basalts and ices). Then, we make a linear combination between them, depending on the percentage of basalts and ices that we define for the solids (in this work we considered three cases, planetesimals composed purely by basalts, composed purely by ices, and composed fifty percent by basalts and fifty percent by ices). We remark here again that in all the cases $Q_{d}^*$ is calculated by Eq.(\ref{eq1-sec2-2-2}), but using the effective radius ($r_{\text{eff}}= 3 (M_{T}+M_{P})/4 \pi \rho$) instead of the target's radius.  

 It should be noted that \citet{LyS2012} also obtain  a  derivation  of  a general  catastrophic
disruption law. However, their work is focus on the gravitational regime, and for bodies of different porosities, while in our model we  adopt  compatible  laws  valid in  the gravitational as well as in the strength regime for non-porous bodies.

\section{Results}
\label{sec3}

We aim to study the impact of the improvements on our planetesimal fragmentation model discussed above on the formation of a giant planet. We carried out a set of different simulations including one phenomenon at a time. Our simulations start at the beginning  of the oligarchic growth  with a  moon-sized embryo located at 5~au from the central star. Initially, the embryo is immersed in an homogeneous single-sized population of planetesimals of 100~km of radius and the disk is ten times more  massive than the Minimum Mass Solar Nebula \citep[MMSN,][]{Hayashi1981}. The simulations stop when the mass  of the  envelope  equals the  core mass, i.e. when the cross-over mass is achieved (in this case we consider that the planet ends its formation in a short period of time after the gaseous runaway growth starts), or at 6 Myr when we consider that the gas of the disk is dissipated. 

\subsection{Baseline model}
\label{sec3-1}

In \cite{Guilera2014} we included a first approach to the modelling of planetesimal fragmentation into
our model of giant planet formation \citep{Guilera2010}, and we studied, for different initial planetesimal sizes and disk masses, how the collisional evolution of the planetesimal population modified the planetary formation process. We remark that in \citet{Guilera2014}, the accretion rates of particles with Stokes number less than the unity, were calculated using the prescription derived by \citet{Inaba2001} in the low-velocity regime. We found that only for initial large planetesimals ($r_{p}=$ 100~km) and massive disks, and only if most of the mass loss in collisions was distributed in larger fragments (see Eq.\ref{eq3-sec2-2}), planetesimal fragmentation favored the relatively rapid formation of a massive core (larger than $10~M_{\oplus}$). If smaller planetesimals are considered, the planetesimal fragmentation process inhibits the formation of massive cores. We remark that these results in agreement with previous works that adopt similar hypothesis for the model of planetesimal fragmentation \citep{Inaba2003,Kobayashi2011, OrmelKobayashi2012}. Thus, we choose, as our baseline  case for this work, the most favorable simulation in \cite{Guilera2014} that corresponds to the case of initial planetesimals of 100~km radius and a 10 MMSN initial disk mass.

\begin{figure}[ht]
  \centering
  \includegraphics[width= 0.49\textwidth]{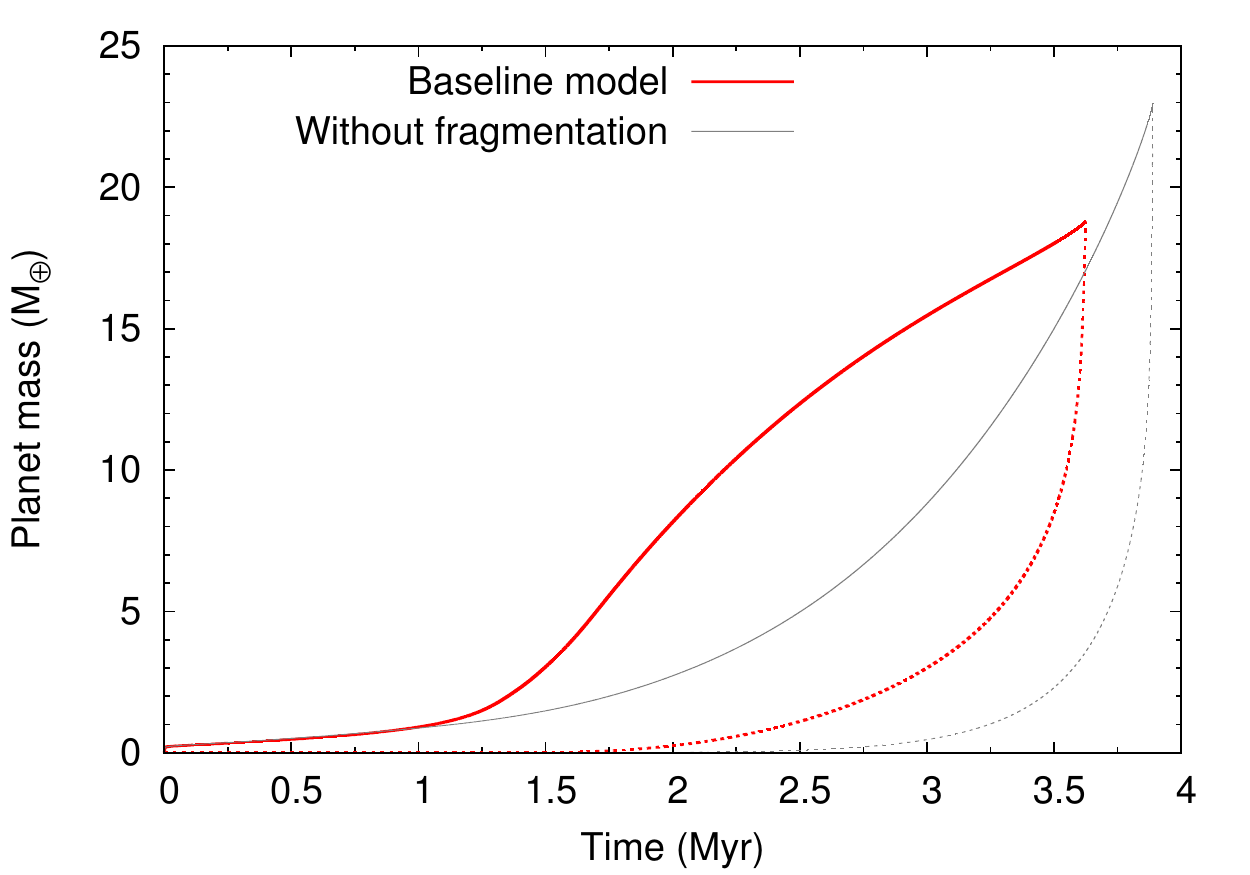}
  \caption{Core masses (solid lines) and envelope masses (dashed lines) as function of time. Here, as in Fig. 5, 8, 9, 12 and 13, the red thick lines  correspond to the baseline model, while  the gray thin lines correspond to the formation of the planet without planetesimal fragmentation. }
 \label{fiducialresult}
\end{figure}

\begin{figure*}[ht]
 \centering
\includegraphics[width= 0.49\textwidth]{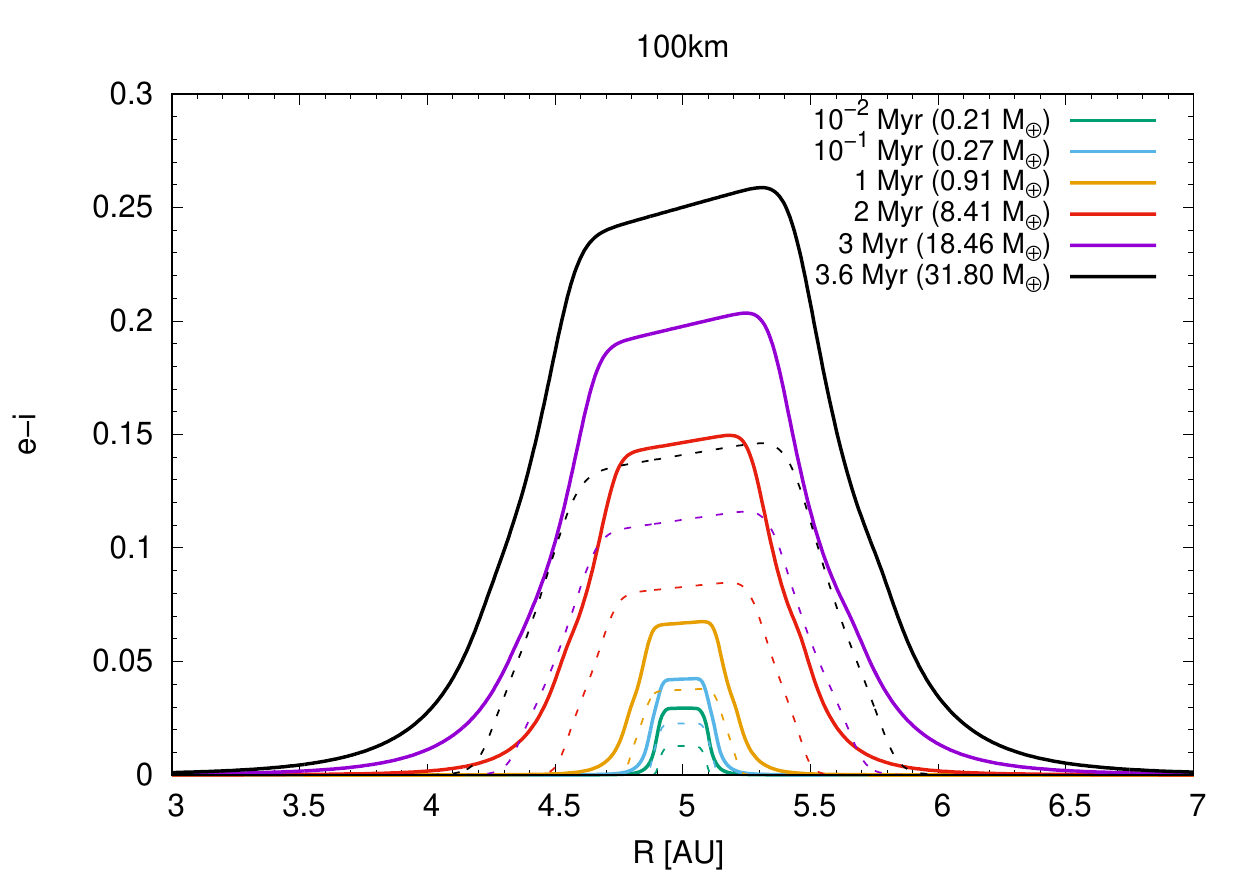}
\includegraphics[width= 0.49\textwidth]{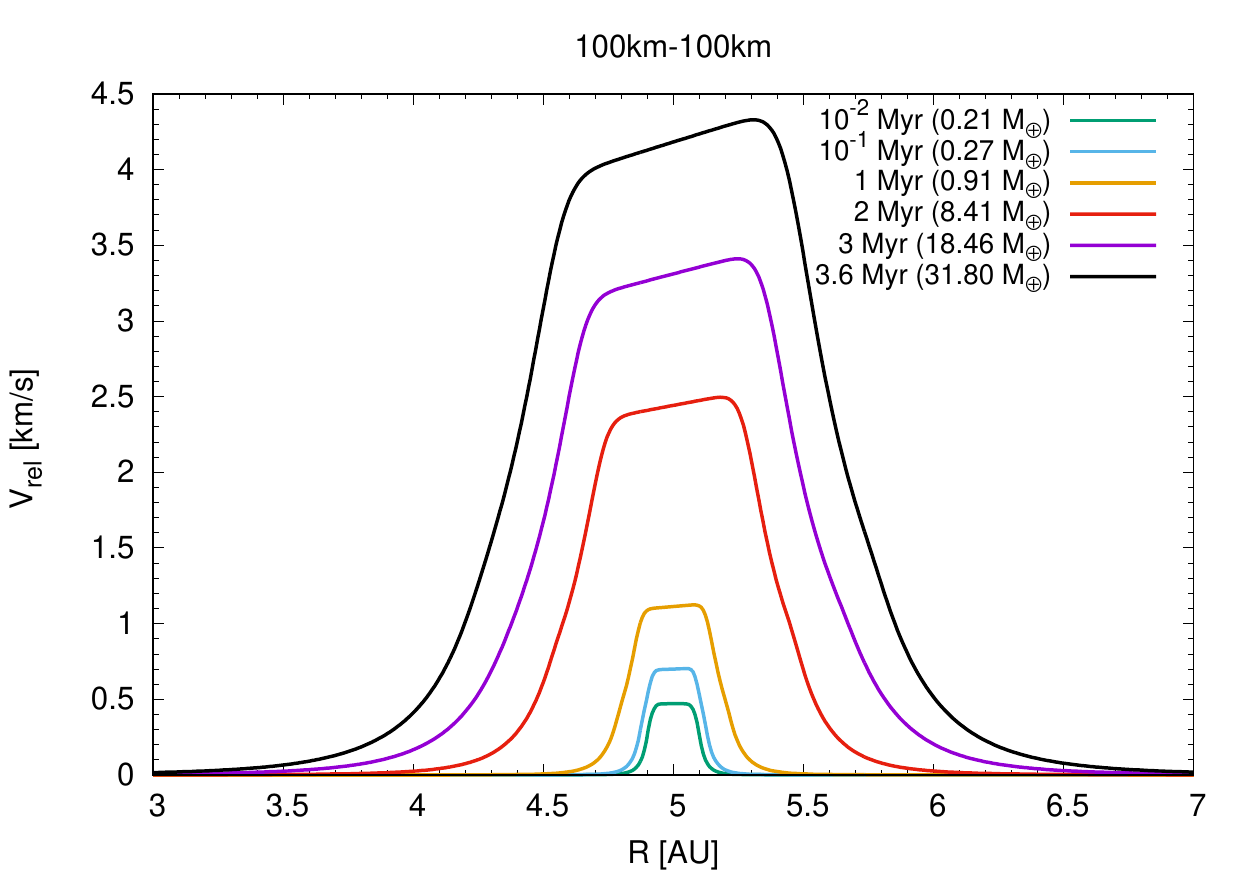}
\caption{Left panel: time evolution of the radial profiles of the eccentricity (solid lines) and inclination (dashed lines) for 100~km sized planetesimals. Right panel: time evolution of the relative velocity between planetesimals with 100~km radius. In brackets we show the total mass of the planet at the given times. The black curve corresponds to the time (and mass)  at which the planet reaches the cross-over mass.}
 \label{fiducialresult2}
\end{figure*}

In Fig.~{\ref{fiducialresult}}, we show the time evolution of the core mass and envelope mass of the planet, for the baseline model and for the formation of the planet without considering the planetesimal fragmentation process. We can see that the inclusion of planetesimal fragmentation reduces by $\sim$ 7\% the time to reach the cross-over mass. We can also observe that the planet reaches the cross-over mass at a lower core mass (18\% lower) . However, the fragmentation model incorporated in \cite{Guilera2014} considered the catastrophic disruption threshold given by \cite{BenzAsphaug1999} for basalts at impact velocities of 3~km/s. This is clearly a simplification. On one hand, as we are studying the formation of a giant planet behind the iceline, it should be taken into account that planetesimals are composed by rocks and ices \citep{Lodders2003}. Also, during the process of planet formation, the planetesimal relative velocities are increased due to the gravitational perturbations produced by the growing planet. In the left panel of Fig.~\ref{fiducialresult2}, we show the increase of eccentricities and inclinations for planetesimals of 100~km of radius, and in the right panel their relative velocities, nearby the planet meanwhile it grows. We can see that eccentricities and inclinations are quickly increased near the planet. This leads to an increase of the planetesimal relative velocities from velocities of some meters per second to velocities of about 5 km/s (right panel of Fig.\ref{fiducialresult2}). These results are showing that clearly the assumption of a $Q_D^*$ given for a constant impact velocity is a simplification in the model of planetesimal fragmentation, and thus, it motivated us to incorporate the following dependencies in the catastrophic disruption threshold.


\subsection{Dependencies of $Q^*_D$ with the planetesimal impact velocities and the compositions}
\label{sec3-2}

\begin{figure}[ht]
 \centering
\includegraphics[width= 0.49\textwidth]{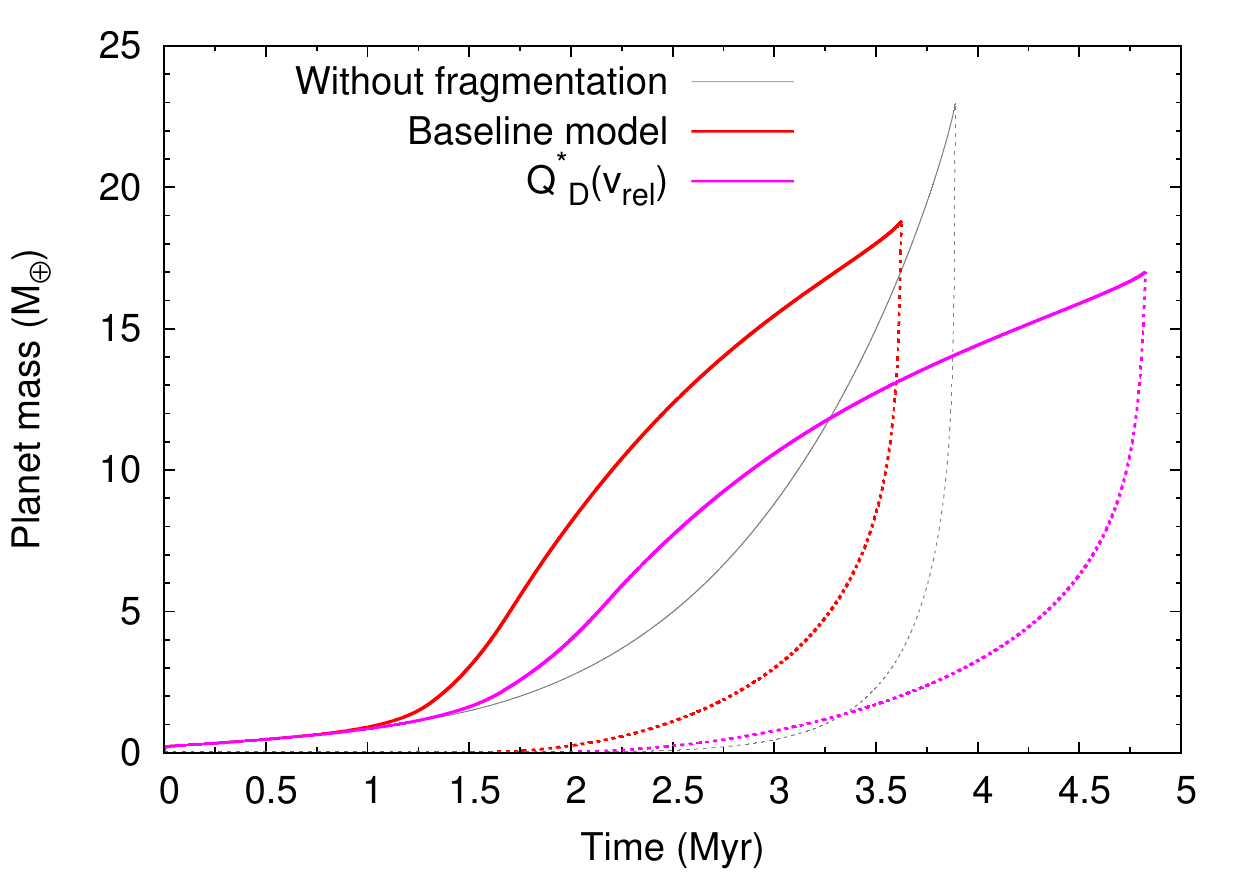}
\caption{Core masses (solid lines) and envelope masses (dashed lines) as a function of the time. Gray lines: model without planetesimal fragmentation. Red lines: baseline model. Pink lines: model in which $Q^*_D$ is a function of relative velocities. }
\label{masacorevstime}
\end{figure}

\begin{figure}[ht]
  \centering
  \includegraphics[width= 0.49\textwidth]{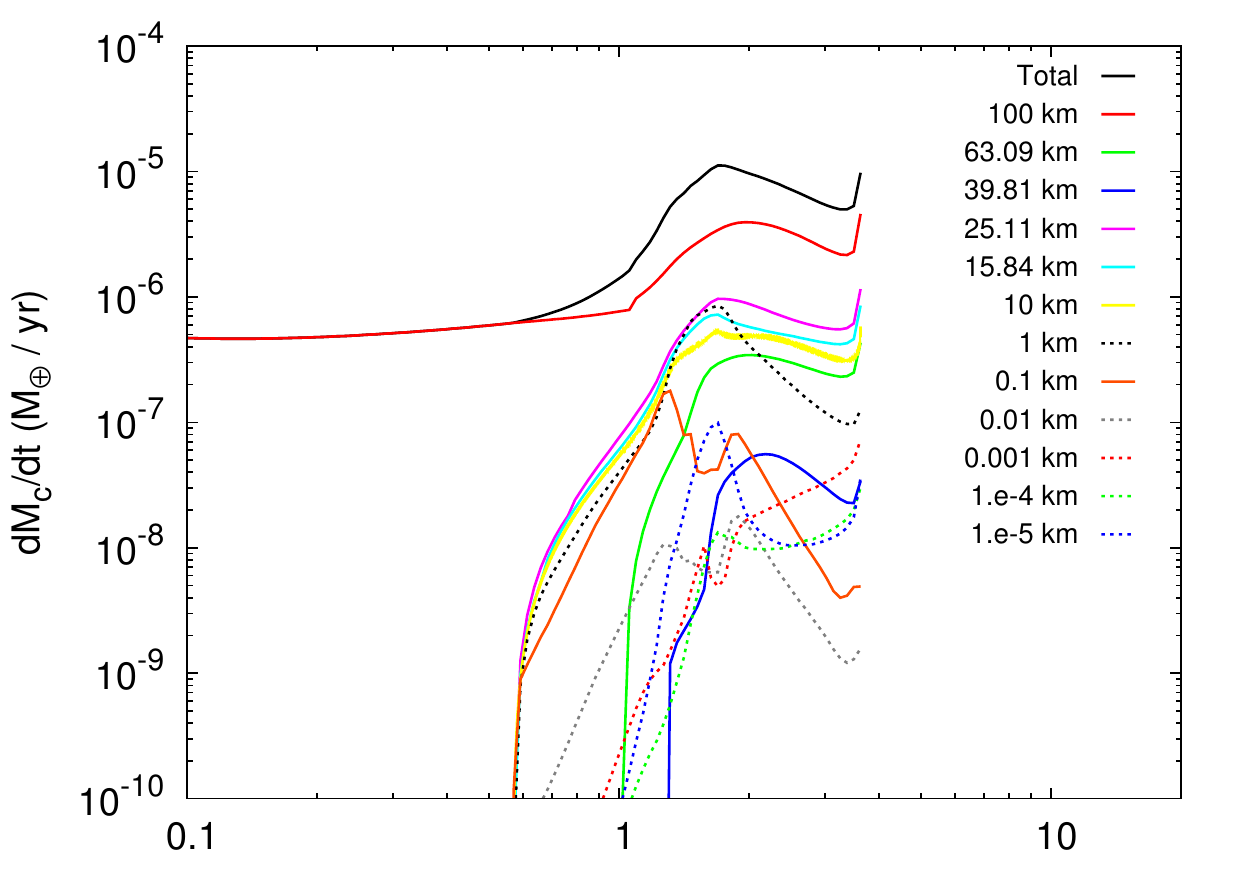} \\
  \includegraphics[width= 0.49\textwidth]{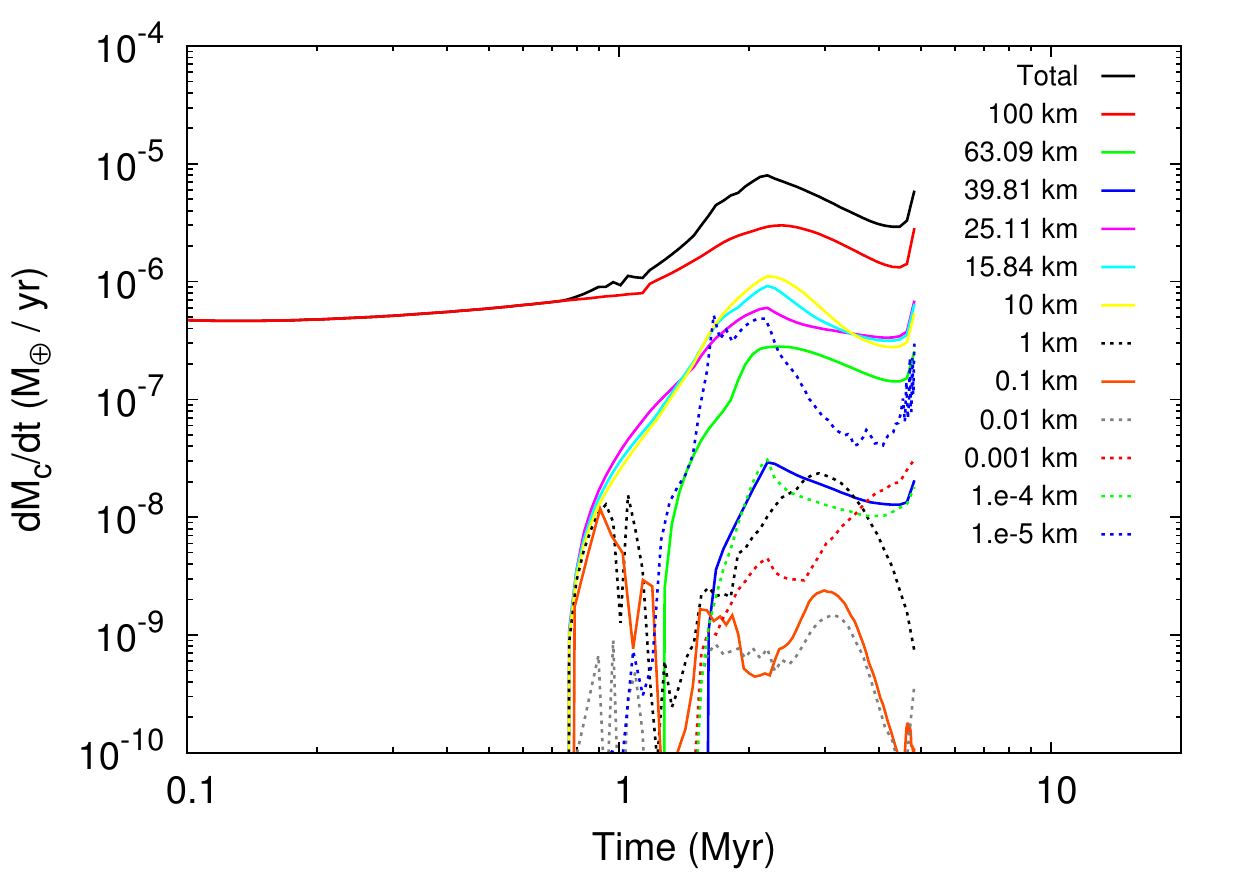} 
  \caption{Top panel: solid accretion rates for basaltic planetesimals and small particles of various radii, products of the collisional evolution of planetesimals, for the baseline model. Bottom panel: same as top panel but for the model that considers $Q^*_D$ as a function of the planetesimal relative velocities.}
  \label{accretionratevstime1}
\end{figure}

\begin{figure}[ht]
 \centering
\includegraphics[width= 0.49\textwidth]{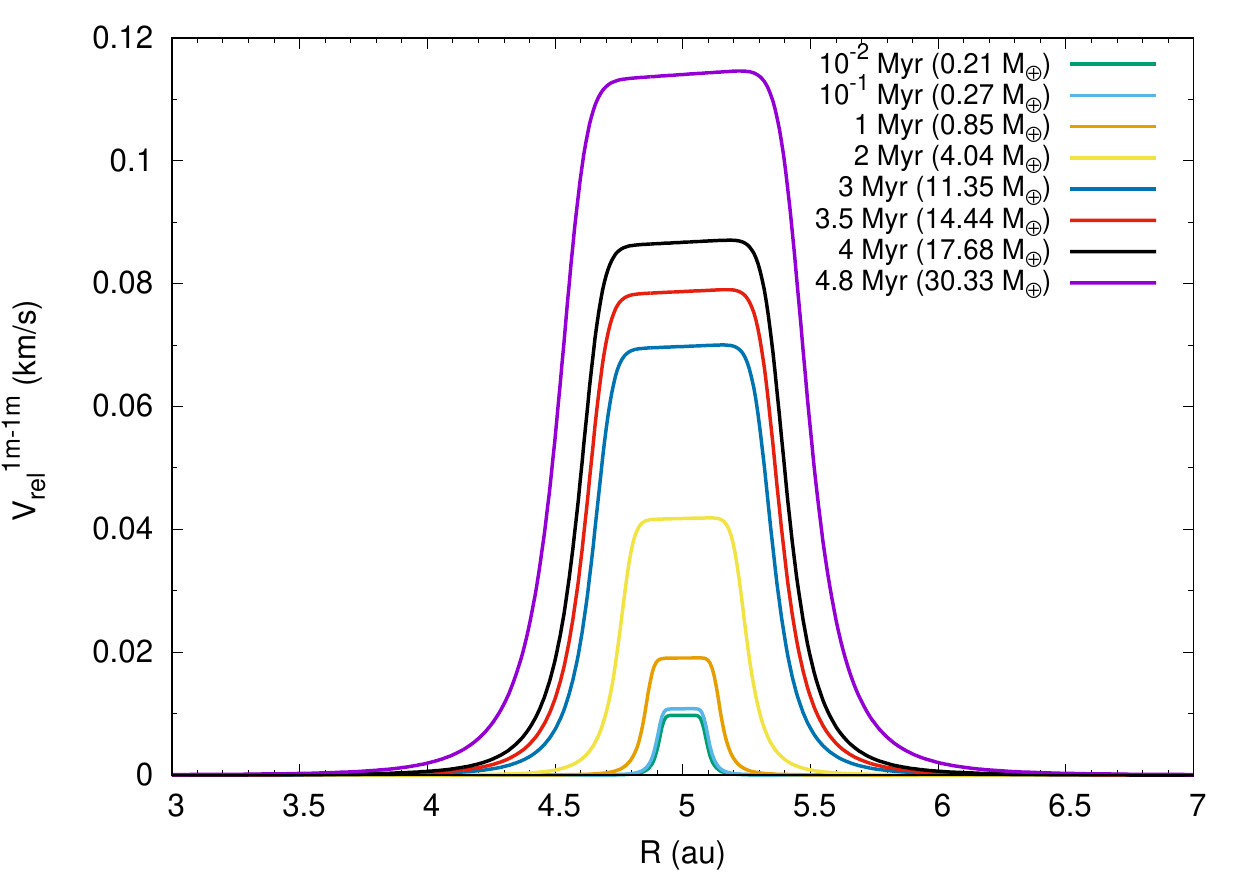}\\
\includegraphics[width= 0.49\textwidth]{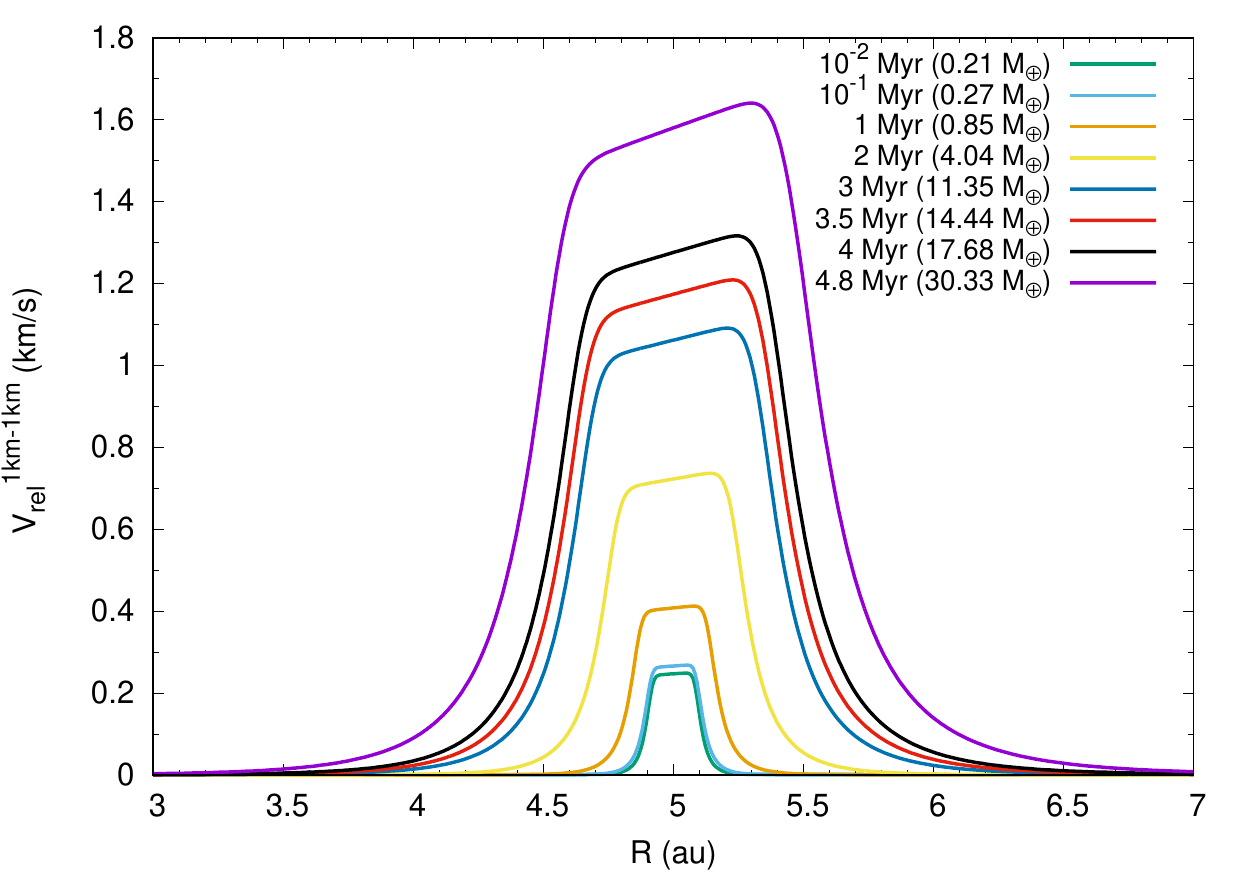}
\caption{ Time evolution of the relative velocity between basaltic planetesimals of the same size for the model that considers $Q_D^*$ as a function of the planetesimal relative velocities. Top panel: 1 m-sized planetesimals. Bottom panel: 1 km-sized planetesimals. In brackets we show the total mass of the planet at the given times. }
 \label{velrel1}
\end{figure}

As we mentioned above, in general, models of planet formation that include planetesimal fragmentation consider the catastrophic impact energy threshold for a fixed velocity of basaltic planetesimals. In this section, we explore how the dependencies of the catastrophic disruption threshold with the planetesimal compositions and with the planetesimal impact relative velocities impact affect the formation of the planet. To do this, we first calculate the formation of the planet considering the same initial parameters as in the baseline model (an embryo located at 5 au in a disk 10 times more massive than the MMSN and with initial planetesimals of 100~km of radius) but now we implement the dependence of $Q_D^*$ with different impact velocities following the approaches described in Sec.~\ref{sec2-2-2}. The solid accretion rates are calculated following Eq.(\ref{eq1-sec2-1}). In Fig.~\ref{masacorevstime}, we show the time evolution of the core mass and envelope mass of the planet, for the model wherein planetesimal fragmentation is not considered, for the baseline model, and for the case where we consider basaltic planetesimals with $Q^*_D$ as a function of the relative velocities among planetesimals. We can see that the cross-over mass, for the last case, is achieved in a time longer than for the baseline model. Moreover, the formation of the giant planet takes $\sim 25$\% more time compared to the case wherein planetesimal fragmentation is not considered. In Fig.~\ref{accretionratevstime1}, we plot the solid accretion rates for the different sizes of planetesimals and small particles for both models, the baseline model and the model that considers $Q^*_D$ as a function of the relative velocities among planetesimals. We can see that the accretion of planetesimals smaller than 100~km and small particles generated by the collisional evolution becomes effective at a longer time for the model where $Q^*_D$ is a function of the relative velocities. We can also see that in both models, the total solid accretion rate is always dominated by plantesimals of 100~km. However, while in the baseline model planetesimals between 1~km and 25~km are the products of the fragmentation process that most contribute to the total solid accretion rate (being the planetesimals of $\sim 25$~km the most important), in the model where $Q^*_D$ is a function of the relative velocities, the most important contribution from the fragments is given by planetesimals between 10~km and $\sim 16$~km. Moreover, in this last case, planetesimals of 1~km has a negligible contribution to the total solid accretion rate. This is due to the fact that these bodies are near the minimum of $Q^*_D$ for low impact velocities, which is for this size, one order of magnitude lower with respect to the $Q^*_D$ at impact velocities of 3~km/s. 
 We can see from Fig. \ref{velrel1} that relative velocities for planetesimals of 1 m reach 20-30 m/s at 1 Myr (when the mass of the core is $\sim 1~\text{M}_{\oplus}$), and 1 km-sized planetesimals exceed these values rapidly, at $10^{-2}$ Myr (when the mass of the core is only $\sim 0.2~\text{M}_{\oplus}$) the relative velocity is over 200 m/s. It is important to remark that, in both cases, their velocities are always below 3~km/s.     

\begin{figure}[ht]
 \centering
\includegraphics[width= 0.49\textwidth]{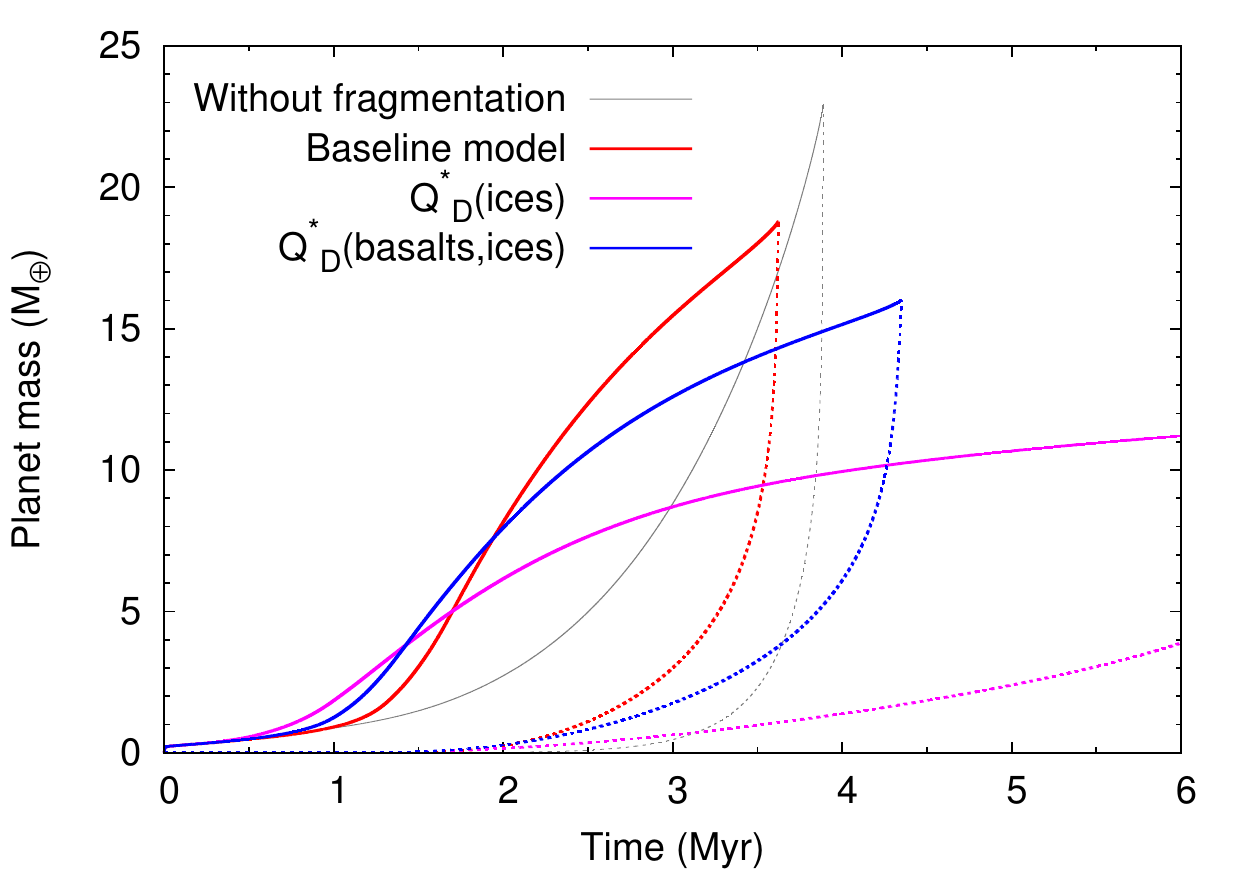}
\caption{ Core masses (solid lines) and envelope masses (dashed lines) as a function of time. Gray lines: model with no planetesimal fragmentation. Red lines: baseline model. Pink lines: model with $Q^*_D$ for ices at impact velocities of 3~km/s. Blue lines: model where $Q^*_D$ is calculated by a linear combination between $Q^*_D$ for basalts at 3~km/s and $Q^*_D$ for ices at 3~km/s. }
\label{masacorevstime2}
\end{figure}

On the other hand, in Fig.~\ref{masacorevstime2} we plot the time evolution of the core mass and envelope mass of the planet, for a model that considers planetesimals purely composed by ices and adopting $Q^*_D$ for ices at impact velocities of 3~km/s, and for a model wherein planetesimals are composed of 50\% of basalts and 50\% of ices and $Q^*_D$ is calculated by a linear combination between $Q^*_D$ for basalts at 3~km/s and $Q^*_D$ for ices at 3~km/s. We also show, by completeness and comparison, the baseline model and the case where planetesimal fragmentation is not considered. We can see that when we considered planetesimals purely composed by ices, the cross-over mass is never achieved. This is due to the fact that for a same fixed impact velocity (in this case 3~km/s), $Q^*_D$ is lower for ices than for basalts. Thus, the collisional evolution of the planetesimal population is different, small fragments are disrupted more efficiently, and the accretion rates of such fragments do not compensate the fragmentation of big bodies.  Besides, if we compare $Q^*_D$ from Fig.\ref{basalts} and Fig.\ref{ice} we can see that  $Q^*_D$ for ices at 3 km/s is lower than  $Q^*_D$ for basalts at 20-30 m/s for planetesimals of $\sim$ 100 km. Therefore, this difference can explain why in the model that considers $Q^*_D$ as a function of the relative velocities for basalts the giant planet core could form during the disk lifetime and, for the model that adopts $Q^*_D$ for ices at 3~km/s, the cross-over mass was not achieved in that time.
 When planetesimals are composed by 50\% of basalts and 50\% of ices, and $Q^*_D$ is calculated as a linear combination of the corresponding $Q^*_D$ for basalts and ices at impact velocities of 3~km/s, respectively, the cross-over mass is achieved in less than 6~Myr with a core mass of 16 $M_{\oplus}$. However, the planet achieves the cross-over mass at a longer time, by $\sim$ 12 $\%$, compared to the case without planetesimal fragmentation.

All these results show that the collisional evolution of the planetesimal population and the accretion of the fragments produced in such collisional evolution play and important role in the formation of a giant planet. We point out that in this subsection, the accretion rates of particles with Stokes number less than the unity, was also computed as in the baseline model, using the prescription derived by \citet{Inaba2001} in the low-velocity regime. 

\subsection{Accretion of small fragments: pebble accretion of second generation}
\label{sec3-3}
 
As we mentioned before, we improved in our model the solid accretion rates incorporating in this work the accretion rates of pebbles (Eq.\ref{eq2-sec2-1}) derived by \citet{Lambrechts2014} for particles with Stokes numbers less or equal to the unity. In \citet{Guilera2014}, the accretion rates of such particles were calculated using the prescription derived by \citet{Inaba2001} in the low-velocity regime (Eq.\ref{eq1-sec2-1}). In that work, we showed that the accretion rates for particles smaller that $\sim 1$~m could become greater than the pebbles accretion rates \citep[see][for the details]{Guilera2014}. Thus, in order to calculate in a more accurate way the accretion of small particles, we implemented the pebble accretion rates mentioned above. 

\begin{figure}[ht]
 \centering
\includegraphics[width= 0.49\textwidth]{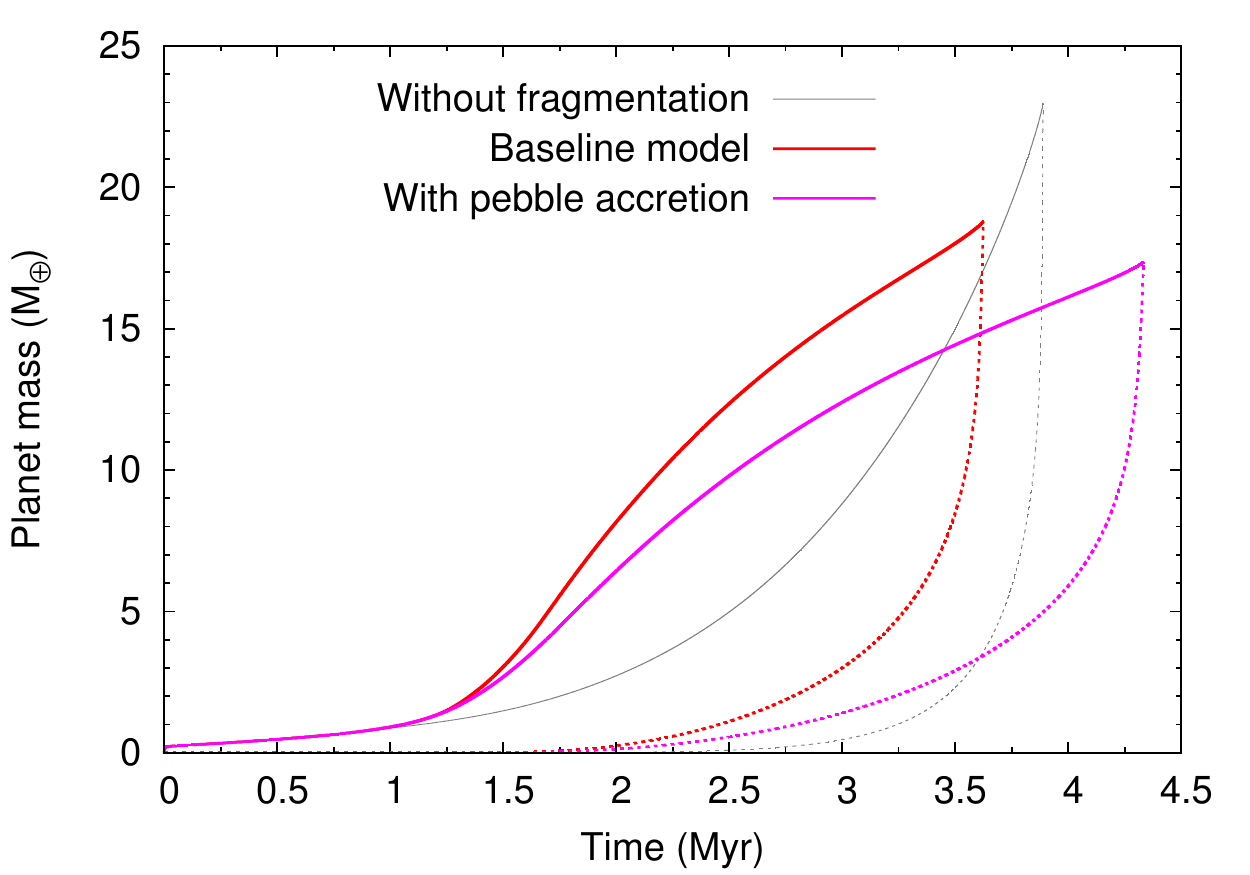}
\caption{ Core masses (solid lines) and envelope masses (dashed lines) as a function of time. Gray lines: model with no planetesimal fragmentation. Red lines: baseline model. Pink lines: model with $Q^*_D$ for basalts at impact velocities of 3~km/s and where small particles with Stokes number less or equal to the unity are accreted as pebbles of second generation.}
\label{masacorevstime3}
\end{figure}

In Fig.~\ref{masacorevstime3}, we show the time evolution of the core mass and the  envelope mass for the case where planetesimals are purely composed by basalts adopting the corresponding $Q^*_D$ at impact velocities of 3~km/s, and where small particles with Stokes number less or equal to the unity are accreted as pebbles using Eq.\ref{eq2-sec2-1}, while in the baseline model the accretion rate for planetesimals (Eq.\ref{eq1-sec2-1}) is applied also for small particles. We also plot for comparison the baseline model and the case wherein planetesimal fragmentation is not considered. We can see that a more accurate treatment for the accretion of small particles delays the formation of the giant planet by about 30\%, despite pebbles do not mostly contribute to the total solid accretion rate (Eq.\ref{eq2-sec2-1}). 

\begin{figure}[ht]
 \centering
\includegraphics[width= 0.49\textwidth]{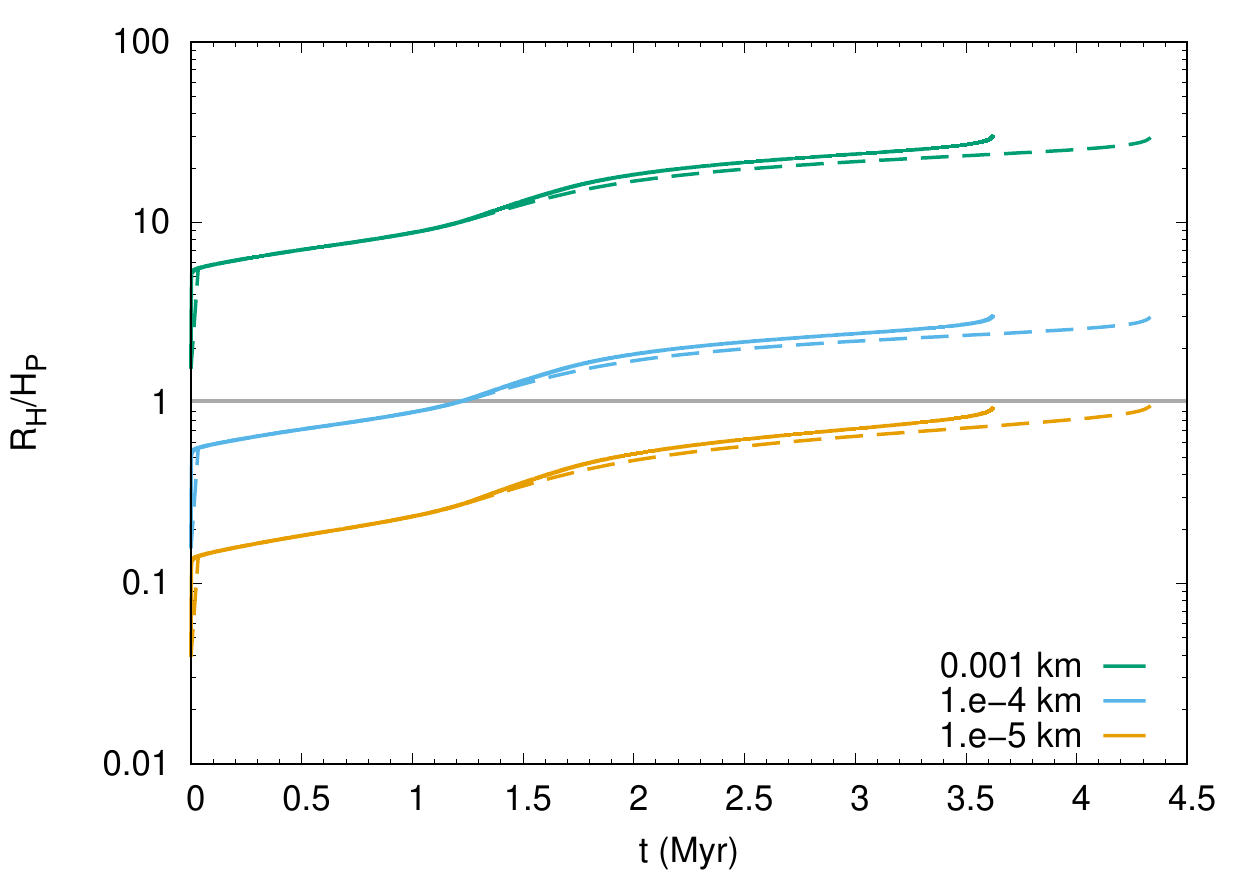}
\caption{ The ratio between the Hill radius and the scale height of small particles, with Stokes number less or equal to the unity, for the baseline model (solid lines) and for the simulation including pebble accretion (dashed lines) as a function of the time. }
\label{scaleheight}
\end{figure}

\begin{figure}[ht]
 \centering
\includegraphics[width= 0.49\textwidth]{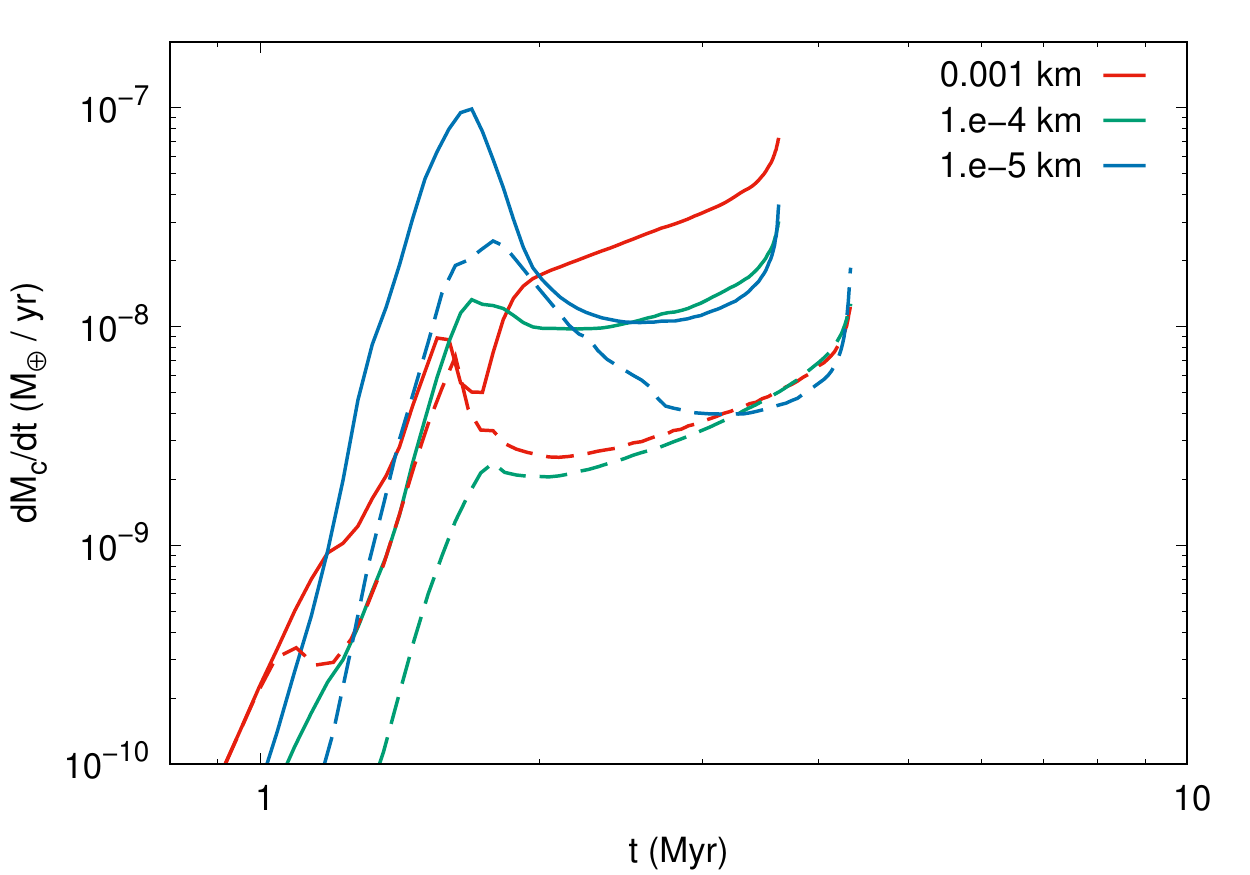}
\caption{Solid accretion rates for small particles, product of the collisional evolution of planetesimals, for the baseline model (solid lines) and for the simulation including pebble accretion (dashed lines) as a function of time. Small particles of these sizes have Stokes number less or equal to the unity. }
\label{accretionratesvstime14}
\end{figure}

This effect is discussed in \citet{Guilera2014} where the rates of small fragments using the prescriptions of \citet{Inaba2001} and \citet{LambrechtsJohansen2012} are compared. Following that analysis we here obtain the accretion rates of \citet{Inaba2001} in terms of the pebble accretion rates of \citet{Lambrechts2014} 

  \begin{eqnarray}
  \dot{M}=
  \begin{cases}
     5.65 \sqrt{\tilde{R}_C/R_H} \dot{M}_H , \quad \text{if } ~ 0.1 \le S_t < 1, \\
    \\
     5.65  \sqrt{\tilde{R}_C/R_H} \left( \frac{S_t}{0.1} \right)^{-2/3} \dot{M}_{H2} , \quad \text{if }~S_t < 0.1 ,
  \end{cases}
  \label{accretionratespebblesplanetesimals}
\end{eqnarray}

\noindent where $\tilde{R}_C$ is the enhanced radius due to the planet’s gaseous envelope, $\dot{M}_H=2 R_H^2 \Sigma_p \Omega_P$ and $\dot{M}_{H2}= \left( \frac{S_t}{0.1} \right)^{2/3} \dot{M}_H$, where $\dot{M}_{H}$ for $0.1 \le S_t < 1$ and $\dot{M}_{H2}$ for $S_t < 0.1$ are the pebble accretion rate $dM_C/dt$ of Eq. \ref{eq2-sec2-1}. We can see from Fig. 15 of \citet{Guilera2014} that when the planet’s core mass becomes larger than $\sim 0.2 M_{\oplus}$ planetesimal accretion rates are higher than pebble accretion rates for fragments with $r_P \lesssim 1$ m. In our work this is the case for particles with $0.1 \le S_t < 1$, but for fragments with  $S_t < 0.1$, the difference between the planetesimal and pebble accretion rates is even higher due to the factor $\left( \frac{S_t}{0.1} \right)^{-2/3}$. 
 Moreover, we introduced a factor $\beta = min (1, R_H/H_P)$ to include a reduction in the pebble accretion rates if the scale height of the pebbles becomes greater than the Hill radius of the planet, this is not taken into account in the planetesimal accretion rate for small fragments in the baseline model. In Fig.~\ref{scaleheight} we show $R_H/H_P$ as a function of time, where  we can observe that objects between $1-10$ cm have values of $R_H/H_P$ below 1. 
 The quantified differences are shown in Fig.~\ref{accretionratesvstime14} where we present the comparison of the solid accretion rates for small particles with $S_t \lesssim 1$ of the baseline case and the simulation including pebble accretion. We can see from Fig.~\ref{accretionratesvstime14} that the accretion rates in the baseline model are higher than in the simulation that takes into account the accretion of pebbles as explained before. 

\subsection{Global model}
\label{sec3-4}

In this section, we compare the results for the formation of the giant planet obtained with the baseline model and with the global model. The global model includes all the improvements on the calculation of $Q^*_D$, i.e., the dependencies of $Q^*_D$ with the planetesimal impact velocities and their compositions described in Section \ref{sec3-2}, and the accretion of second generation pebbles described in Section \ref{sec3-3}. 

\begin{figure}[ht]
  \centering
 \includegraphics[width= 0.49\textwidth]{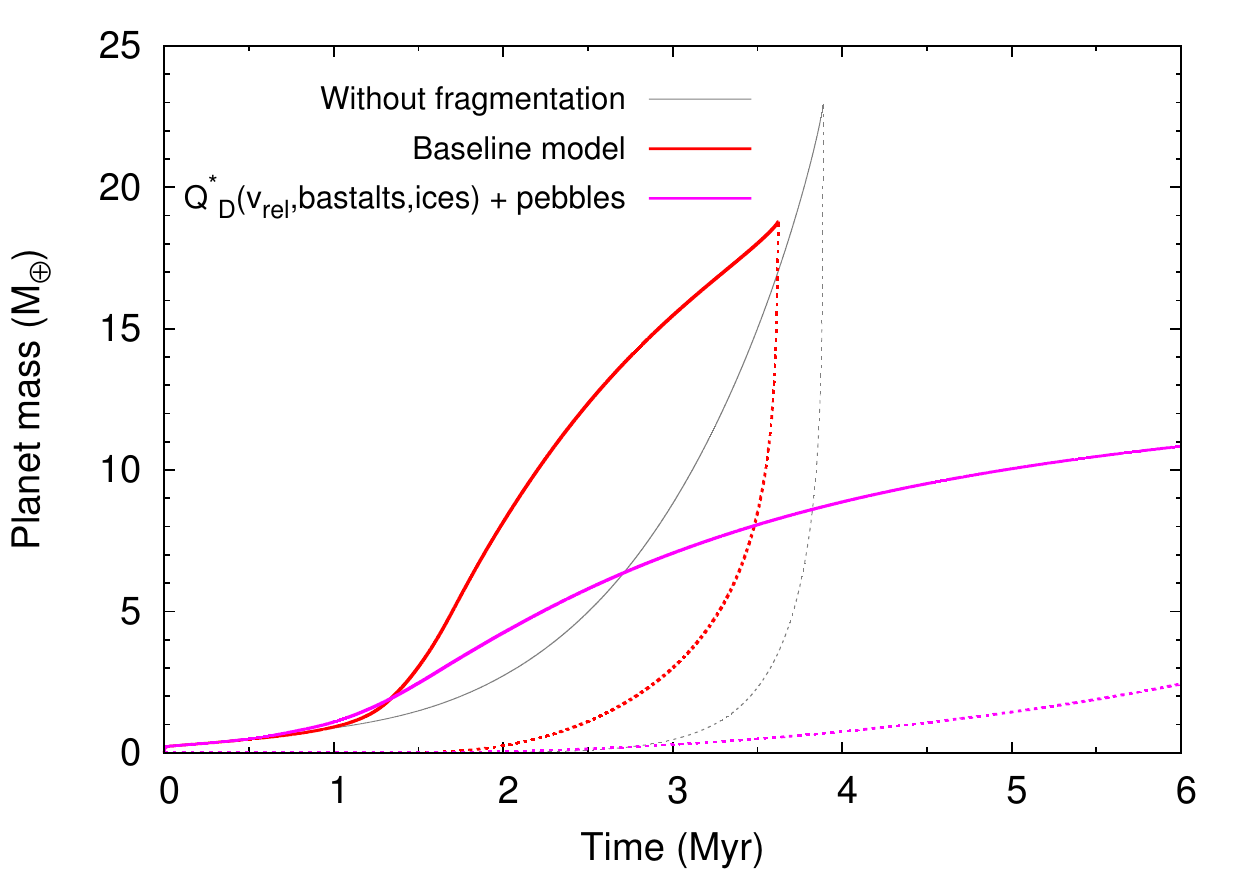}
 \caption{ Core masses (solid lines) and envelope masses (dashed lines) as a function of time. Gray lines: model without planetesimal fragmentation. Red lines: baseline model. Pink lines: model with all the improvements on the calculation of $Q^*_D$ (the dependencies of $Q^*_D$ with the planetesimal impact velocities and compositions) and the pebble accretion rates.}
 \label{masacorevstime4}
\end{figure}

In the global model where we include all the new effects on $Q^*_D$ and a more accurate and realistic treatment for the accretion of the small particles, the core does not achieve the cross-over mass within the 6 Myr as shown in Fig.~\ref{masacorevstime4}. We point out this model also includes the different regimes for the calculation of the relative velocities (and the impact rates) discussed in Sec.~\ref{sec-2-2-1}. However, as the planet quickly excites the relative velocities of the near planetesimals, they are in general in the dispersion regime, and thus the keplerian shear regime does not play a relevant role. Finally, we remark that despite the planet does not achieve the cross-over mass, a solid core of a few Earth masses, up to 5 Earth masses, is formed more quickly compared to the case where planetesimal fragmentation is not considered. \citet{Ikoma2000} and \citet{Hubickyj2005} showed that a reduction in the grain opacity in the planet envelope as well as the pollution of the envelope \citep[due to evaporated materials of icy planetesimals in the envelope, see][]{Hori-Ikoma-2011}, could reduce the mass of the core at which the planet reaches the cross-over mass, and thus it reduces the formation time. We will explore this possibility in the next section.  

 \subsection{Reduced grain opacities}
\label{sec3-5}

 In the previous sections we studied the process of giant planet formation considering fragmentation
of plantesimals including for the catastrophic disruption threshold the dependencies with relative velocity and composition of the colliding  bodies. We also
include the accretion rates for small fragments called pebbles. We analyzed every improvement separately and finally all together and we found that, in all the cases, the formation of the giant planet core is slower.

Motivated by this result we incorporate a physical phenomenon that could act in an opposite way, accelerating the formation of a giant planet, in this case a reduction in the grain opacities of the planet envelope.

The grain opacities in the planet envelope play an important role on the formation of a giant planet. \cite{Ikoma2000} found that if the grain opacities are small enough, a small size giant planet core can capture significant amounts of gas. Later, \cite{Hubickyj2005} found that if the grain opacities are assumed as 2\% of the interstellar medium value, the planet can reach the cross-over mass with a core mass between $5-10~M_{\oplus}$. More recently, \citet{Hori-Ikoma-2010} showed that a core of only $\sim 2~M_{\oplus}$ is able to capture enough gas to form a giant planet if the accreting envelope is grain-free. Thus, following these works, specially the work of \citet{Hubickyj2005}, we compute a set of simulations, with and without planetesimal fragmentation, reducing the grain opacities in the envelope up to a 2\% of the interstellar medium value. For the case where we consider the planetesimal fragmentation, we adopt the global model described above. 

\begin{figure}[ht]
 \centering
\includegraphics[width= 0.49\textwidth]{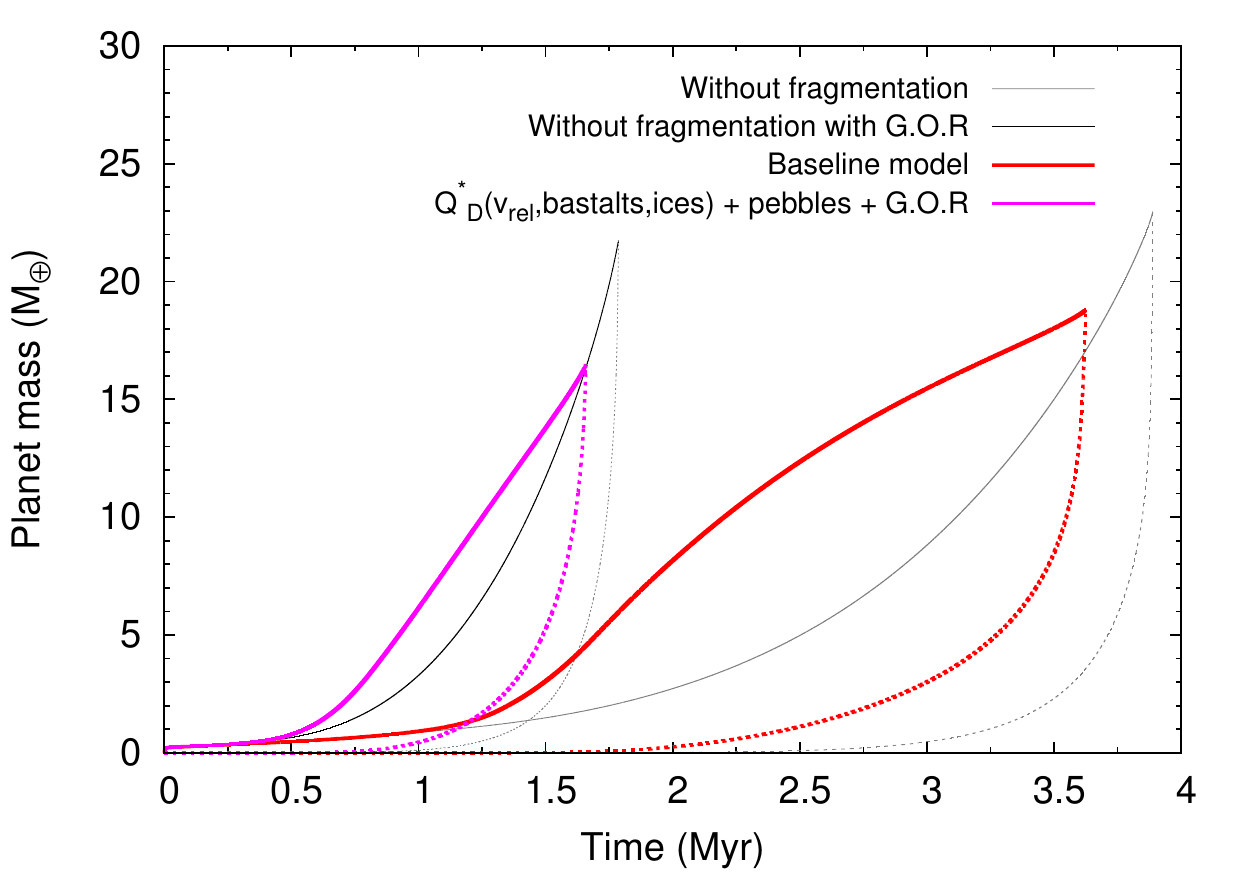}
\caption{ Core masses (solid lines) and envelope masses (dashed lines) as a function of time. Gray lines: model without planetesimal fragmentation. Black lines: model without planetesimal fragmentation but considering grain opacities reduction (G.O.R). Red lines: baseline model. Pink lines: model taken into account all the improvements on the calculation of $Q^*_D$ (the dependencies of $Q^*_D$ with the planetesimal impact velocities and compositions), pebble accretion rates and considering reduced grain opacities. }
 \label{masacorevstime5}
\end{figure}

\begin{figure}[ht]
 \centering
\includegraphics[width= 0.49\textwidth]{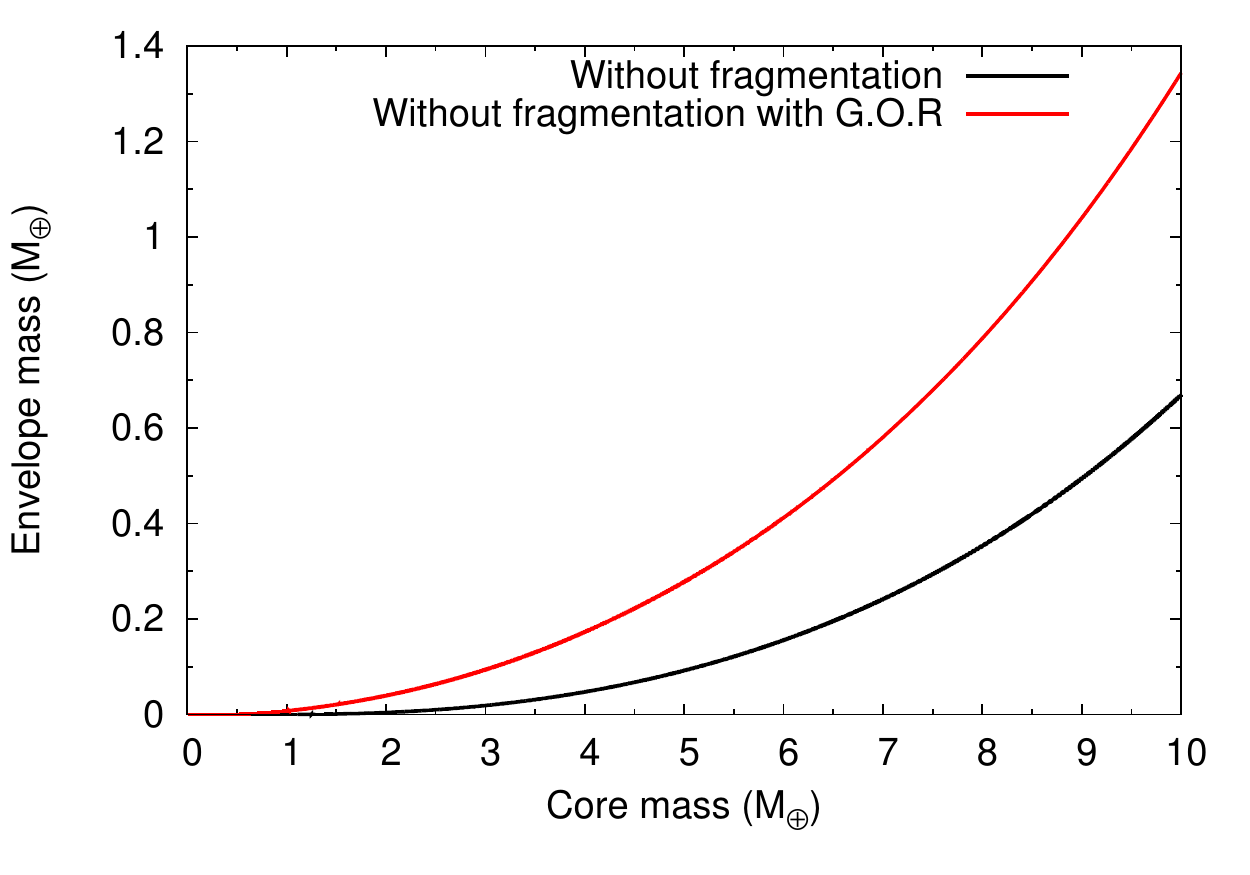}
\caption{Envelope mass as a function of the core mass for the two cases where planetesimal fragmentation is not considered. We plot the core mass until $10~M_{\oplus}$ in order to show more clearly the differences at low-mass core values. The black line represents the case where grain opacities are not reduced, while the red line represents the case where grain opacities are reduced in the planet envelope.}
 \label{mgas-vs-mcore}
\end{figure}

In Fig.~\ref{masacorevstime5}, we present the time evolution of the core mass and the envelope mass of the planet for the models that consider reduced grain pacities with and without planetesimal fragmentation and the baseline case. When we calculate the formation of the giant planet without considering the planetesimal fragmentation process, the time at which the planet achieves the cross-over mass is reduced in more than 50\% when the grain opacities in the planet envelope are reduced. However, the cross-over mass remains practically similar. These results are in very good agreement with the previous results of \citet{Hubickyj2005}. A reduction in the envelope opacity allows the planet to release more efficiently the heat generated by the accretion of the planetesimals, and as consequence, the gas accretion becomes more efficient. We can see this effect in Fig.\ref{mgas-vs-mcore} where we plot the envelope mass as a function of the core mass for the two models without planetesimal fragmentation. At a fixed core mass, the model with the reduced grain opacities has a greater value for the envelope mass. 

Finally, we note that when planetesimal fragmentation is considered, the planet achieves the cross-over mass at a time (1.66 Myr) and with a core mass (16.44 $M_{\oplus}$) that are lower than for the case where planetesimal fragmentation is not considered ($\sim$ 7$\%$ and $\sim$ 24$\%$ lower in time and core mass respectively). We associate this result to the fact that the amount of envelope mass in the planet at low-mass cores plays an important role, significantly enhancing the capture radius of the planet.

\vspace{0.5 cm}

\section{Comparison with previous work}
\label{sec3-6}

A similar model of giant planet formation was developed by \citet{Chambers2014}. They carried out numerical simulations of oligarchic growth including pebble accretion and planetesimal fragmentation. One of the major differences between their model and ours is that they consider that all the collisions involving targets smaller than their minimum bin size are assumed to coagulate and grow rapidly approaching such minimum size, while we assume that the mass distributed under $r_p^{min}=1$ cm is lost. In our fragmentation model most of the mass is in the largest fragments. As discussed in \citet{Guilera2014}, the results are analogous when the minimum bin size $r_p^{min}$ is reduced to lower values.
Also, \citet{Chambers2014} uses a fixed value of $Q_D^*$ (for basalts at 3 km/s) and includes an additional factor to reduce 10 and 100 times the strength for all the sizes of the bodies to see how this could change the growth of the embryo. He found the most significant differences in the case where $Q_D^*$ is reduced by a factor of 100, changing the growth track of the embryo (shown in Fig. 20 of his paper). Even though we can't compare directly this results with ours since our models are different, we agree that, the relative importance of pebble accretion for giant planet formation will depend strongly on the rate at which pebbles are generated from plantesimal-planetesimal collisions, depending in particular on the impact parameters.

Another model of giant planet formation that includes pebble accretion is the one of \citet{Alibert2018}. They  developed a model of the formation of Jupiter that could explain the constrains that the cosmochemical evidence gives about the existence of two main reservoirs of small bodies that remain separated during the early Solar System \citep{Kruijer2017}. They conclude that Jupiter formed in a three step process. First, Jupiter's core grew by the accretion of pebbles. Second, the pebble accretion stopped and the core continued growing by the accretion of small planetesimals. Third, the core was massive enough for the runaway gas accretion to start. It is important to highlight that our initial conditions correspond to the beginning of the oligarchic growth where almost all the solid material is contained in planetesimals, in contrast to \citet{Alibert2018} where they start their simulations with a first generation of pebbles. Moreover, they don't include a fragmentation model since they estimate, for the start of their second stage, the amount of fragments that could have been produced during their first stage of pebble accretion. For this estimation, they used a fixed value of $Q_D^*$ for 100 km-sized planetesimals. Different initial conditions for giant planet formation including a first generation of pebbles and a variation of the initial size of planetesimals will be studied in future works.

\section{Summary and conclusions}
\label{sec4}

We studied the formation of a giant planet considering the collisional evolution of the planetesimal population, including the dependence of the catastrophic impact energy threshold with composition and relative velocities of the colliding planetesimals. We also included the pebble accretion rates for small particles product of the collisional cascade which are pebbles of second generation strongly coupled to the gas to calculate in a more accurate way than in our baseline case.  We showed that as the planet grows, the relative velocities among the planetesimals near the planet are increased from velocities of the order of m/s to velocities up to $\sim 4.5$~km/s before the planet reaches the cross-over mass.

 We analyzed the improvements incorporated in the calculation of $Q_{D}^*$  and the inclusion of the pebble accretion rates one at each time, and finally all the improvements together. We point out that, for each improvement in $Q_{D}^*$, as the collisional cascade increases, the timescale to reach the cross-over mass is delayed.  We also found that in all the cases, the formation  of the  core is less efficient compared with the case where planetesimal fragmentation is not considered, and with our baseline model.

 When we include all the improvements at once there is a sum of effects that inhibit the formation of the giant planet core.   On one hand, pebble accretion rates are lower than the plantesimal accretion rates for the small particles adopted in the baseline model. This effect contributes to slow down the growth of the giant planet core.   On the other hand, when we include the dependence of $Q_{D}^*$ with impact velocity and composition we first interpolate between the curves of $Q_{D}^*$ for a given velocity obtaining $Q_{D}^*$ for each pure material. Then, we make a linear combination between these values where we assume that planetesimals are half basalts and half ices. For low impact velocities basalts are weaker than than ice (except for $\sim$ 100 km-sized bodies) and for higher velocities, ice is weaker than basalts. These effects are averaged since the composition of planetesimals are 50\% basalts and 50\% ices. Moreover, it is important to remark that the catastrophic disruption threshold for ices at impact velocities of 0.5 km/s and 3 km/s are lower for $\sim$ 100 km-sized plantesimals than $Q_{D}^*$ for basalts at any of the three impact velocities (20-30 m/s, 3 km/s and 5 km/s) for that size of bodies.

 Regardless of whether the planet arrives to the cross-over mass or not within the disk lifetime, in all the cases, planetesimal fragmentation favors the relative rapid growth of the giant planet core up to a few Earth masses compared to the case where plantesimal fragmentation is not considered. 

This last result motivated us to explore alternatives for the formation of the giant planet with less massive cores. In this way, we followed the work of \citet{Hubickyj2005}. These authors showed that if the grain opacity of the planet envelope is reduced up to a 2\% of the interstellar medium value, the planet can reach the cross-over mass with a core mass between $5-10~M_{\oplus}$. Thus we computed two new simulations, reducing the grain opacity of the envelope up to a 2\% of the interstellar medium value, considering planetesimal fragmentation (including all the improvements in our model) and without planetesimal fragmentation. For this last case, we found that the planet reached the cross-over mass at a time significantly lower (by a factor two) with respect to the case without planetesimal fragmentation but not considering reduced grain opacities. However, the mass of the core at which the cross-over mass is achieved remained practically similar. For the case where planetesimal fragmentation and reduced grain opacities are included, the planet reached the cross-over time at a lower time and with a less massive core compared to the case where reduced grain opacities  are included but without planetesimal fragmentation. We associated this result to the fact that if the opacity of the planet envelope is reduced, the planet release more efficiently the luminosity through the envelope, and then, the gas accretion becomes more efficient. Thus the planet accretes significantly more gas for a low-mass core, enhancing the radius of capture of the planet and increasing the solid accretion rate.   

\citet{Hori-Ikoma-2011} studied the formation of a giant planet where the gas envelope is polluted by icy planetesimals that are dissolved in the envelope. They found that the increase in the molecular weight and the reduction in the adiabatic temperature gradient produced by the pollution of planetesimals, significantly reduce the mass of the core at which the planet achieves the cross-over mass. More recently, \citet{Venturini2015, Venturini2016} found similar results computing for the first time, self-consistent models of giant planet formation that include the effect of the envelope enrichment by the pollution of planetesimals dissolved as they income the planet envelope. The increment in the envelope molecular weight and the reduction in adiabatic temperature gradient produce that the envelope layers are compressed more efficiently and the gas accretion is significantly increased. We will incorporate this phenomenon in a future work. 

\begin{acknowledgements}  
We would like to thank John Chambers and an anonymous referee for the comments and suggestions that helped us to improve this paper. This work was supported by the PIP 112-201501-00699CO grant from Consejo Nacional  de Investigaciones Cient\'{\i}ficas y  T\'ecnicas, and by PIDT 11/G144 grant from Universidad Nacional de La Plata, Argentina. O.M.G is also supported by the PICT 2016-0023 from ANPCyT, Argentina. O.M.G. also acknowledges the hosting as invited researcher from IA-PUC. 

\end{acknowledgements}

\bibliographystyle{aa}

\bibliography{bibliografia}

\begin{appendix}
 \section{Accretion model}
\label{sec5}

\subsection{Evolution of the envelope and planetesimal radial migration}
\label{sec5-1}

 The evolution of the envelope is calculated solving the standard equations  of  the stellar evolution theory. The following equations correspond to the conservation of mass, the hydrostatic equilibrium, the energetic balance and the energy transport 

\begin{equation}
\frac{\partial r}{\partial m_r} = \frac{1}{4 \pi r^2 \rho},   
\end{equation}

\begin{equation}
\frac{\partial P}{\partial m_r} = -\frac{G m_r}{4 \pi r^4},   
\end{equation}

\begin{equation}
\frac{\partial L_r}{\partial m_r} = \epsilon_{pl} - T \frac{\partial S}{\partial t} ,  
\end{equation}

\begin{equation}
\frac{\partial T}{\partial m_r} = - \frac{G m_r T }{4 \pi r^4 P} \nabla ,   
\end{equation}

\noindent where $G$ is the universal gravitational constant,  $\rho$ is the density of the envelope,  $S$ is the entropy per unit mass, $\epsilon_{pl}$ is the energy release rate due to the accretion
of planetesimals, and $\nabla \equiv \frac{d ln T}{d ln P}$ is the dimensionless temperature gradient, which depends on whether the energy is carried by radiation or by convection \citep[see][for details]{Fortier2009,Guilera2010}.

 In our simulations we consider 2500 radial bins logarithmically equally spaced in a protoplanetary disk defined between 0.4 AU and 20 AU. The particles migrate inward in the protoplanetary disk due to gas drag, where the radial migration velocity for each drag regime is given by

\begin{equation}
v_{mig} =
  \begin{cases}
    - \frac{2 a \eta}{t_{stop}}  \left[ \frac{S_t^2}{1+S_t^2}\right]   \quad \text{ Epstein regime}  \\ \\
    -\frac{2 a \eta}{t_{stop}} \left[ \frac{S_t^2}{1+S_t^2}\right]  \quad \text{ Stokes regime} \\ \\ 
    -\frac{2 a \eta}{t_{stop}}  \;\;\;\;\;\;\;\;\;\; \quad \text{ quadratic regime},  
  \end{cases}
\label{eq1-sec2}
\end{equation}

\noindent being $a$ the semimajor axis, $\eta$ the ratio of the gas velocity to the local Keplerian velocity, $S_t= t_{stop} \omega_k$  the Stokes number,  with $t_{stop}$ the stopping time depending on the drag regime (Epstein, Stokes or quadratic regime) and $\omega_k$ the Keplerian frequency. 
 The rest of the model is explained in detail in \citet{Guilera2010, Guilera2014}.

\subsection{Solid accretion rates}
\label{sec5-2}

 For planetesimals, we use the accretion rates given by \cite{Inaba2001}
\begin{eqnarray}
\frac{dM_C}{dt}= R_H^2 \Sigma_P \Omega_{P} P_{\text{coll}}, \;\;\; \text{when} \;\;\; S_{t} \geq 1,
\label{eq1-sec2-1}
\end{eqnarray}
where $M_{C}$ is the core's mass, $R_H$ is the planet's Hill radius, $\Sigma_P$ is the surface density of solids at the location of the planet, $\Omega_{P}$ is the Keplerian frequency at the planet location, and $P_{\text{coll}}$ is the collision probability, which is a function of the core radius $R_C$, the Hill radius of the planet, and the relative velocity between the planetesimals and the planet $v_{\text{rel}}$, thus $P_{\text{coll}}=P_{\text{coll}}(R_C,R_H, v_{\text{rel}})$. In fact, as we also consider the drag force that planetesimals experience on entering the planetary envelope \citep[following][]{Inaba&Ikoma-2003}, the collision probability is function of the enhanced radius $\tilde{R}_C$ instead of $R_C$ \citep{Guilera2014}.

For the pebbles of second generation, we use the pebble accretion rates given by \cite{Lambrechts2014}
\begin{eqnarray}
  \frac{dM_C}{dt} =
  \begin{cases}
     2 \beta R_H^2\Sigma_p \Omega_P, \quad \text{if } ~ 0.1 \le S_t < 1, \\
    \\
     2 \beta \left( \frac{S_t}{0.1} \right)^{2/3} R_H^2\Sigma_p \Omega_P, \quad \text{if }~S_t < 0.1.
  \end{cases}
  \label{eq2-sec2-1}
\end{eqnarray}
In Eq.~(\ref{eq2-sec2-1}), we introduce the factor $\beta= \text{min}(1, R_H/H_p)$ in order to take into account a reduction in the pebble accretion rates if the scale height of the pebbles, $H_p$, becomes greater than the Hill radius of the planet. This can happen if the vertical turbulent dispersion of small particles become significant. The scale height of the solids at a given distance from the central star is given by \citep{Youdin2007}
\begin{eqnarray}
  H_p= H_g\sqrt{\frac{\alpha}{\alpha + S_t}},
  \label{eq3-sec2-1}
\end{eqnarray}
 being $H_g$ the scale height of the disk, and $\alpha$ the dimensionless Shakura \& Sunyaev viscosity-parameter \citep{Shakura-Sunyaev1973} we adopt $\alpha = 10^{-3}$.

\section{Fragmentation model}
\label{sec6}

 According to our model, a collision between a target of mass $M_{T}$ and a projectile of mass $M_{P}$ results in a remnant of mass $M_{R}$ given by 
\begin{equation}
 M_{R} =
  \begin{cases}
    \left[ -\frac{1}{2} \left( \frac{Q}{Q_{D}^*}- 1 \right) + \frac{1}{2} \right] (M_{T} + M_{P}),     & \quad \text{if } Q < Q_{D}*, \\ \\
    \left[ -0.35 \left( \frac{Q}{Q_{D}^*}- 1 \right) + \frac{1}{2} \right] (M_{T} + M_{P}), & \quad \text{if } Q > Q_{D}*, 
  \end{cases}
\label{eq1-sec2-2}
\end{equation}
where $Q$ is the collisional energy per unit target mass and $Q_{D}^*$ is the catastrophic impact energy threshold per unit target mass required to fragment and disperse half of the target mass. Usually, $Q_{D}^*$ is a function of the target's radius. However, it is important to remark here that in our model \citep[following][]{Morbidelli2009} $Q_{D}^*$ has to be calculated with an effective radius $r_{\text{eff}}= 3 (M_{T}+M_{P})/4 \pi \rho$, being $\rho$ the planetesimal density. Besides this, in our model the mass loss in the collision, which we define as $(M_T + M_P - M_R)$, is distributed between the minimum mass bin considered and the mass bin corresponding to the biggest fragment $M_F$ given by 
\begin{eqnarray}
M_F= 8 \times 10^{-3} \left[ \frac{Q}{Q_D^*}~e^{-(Q/4Q_D^*)^2} \right]~(M_T + M_P).
\label{eq2-sec2-2}
\end{eqnarray}
As in \citet{Guilera2014}, we note again that for some supercatastrophic collisions (which occur when $M_R \ll M_T+M_P$),  $M_F > M_R$. For such collisions we assume that $M_F= 0.5 M_R$.

Unlike \cite{Morbidelli2009}, the fragments are distributed following a power-law distribution given by \citep{Kobayashi2011, OrmelKobayashi2012} 
\begin{equation}
\frac{dn}{dm} \propto m^{-5/3},
\label{eq3-sec2-2}
\end{equation}
meaning that most of the mass is distributed in the larger fragments. In \cite{Guilera2014}, we found that  massive cores formation is favored when the exponent of the power-law mass distribution is lower than 2, thus in this work we analyze the most favorable scenarios, taking the value of 5/3 for the exponent. To calculate the growth of the planet, we adopt a discrete planetesimal size distribution using 36 size (or mass) bins logarithmic equally spaced between 1~cm and 100~km, where initially all the solid mass is in the non porous planetesimals of 100~km of radius. However, when we calculate the planetesimal fragmentation process, we extrapolate the planetesimal size distribution two orders of magnitude below the minimum size $r_p^{min}$ of the main model to avoid the accumulation of spurious mass in the smaller fragments. Therefore, only the mass ejected from the collision distributed between the mass of the larger fragment and the minimum size considered  ($r_p^{min}=1$~cm) is taken into account to calculate later the solid accretions rates, i.e., we assume that the mass distributed below 1~cm is lost. 

 The feeding zone of the embryo extends to four Hill radii at either side of the embryo. Adopting 2500 radial bins along the protoplanetary disk guarantees that there are at least ten radial bins between $R_P - 4 R_H$ and $R_P + 4 R_H$ at the beginning of the simulation. We define the width of the fragmentation zone as twice the feeding zone, i.e., eight times the embryo Hill radius at both sides of the embryo. The excitation of eccentricities and inclinations decay with the distance to the embryo, specially outside the feeding zone. Then, our definition of the fragmentation zone guarantees that collisions are well determined within this zone.
The amount of bins increase as the core mass grows,  e.g., in the baseline case (see Sec. \ref{sec3-1}), when the planet reaches the cross-over mass, the fragmentation zone has $\sim 600$ radial bins.

\subsection{Velocities and probabilities of collision regimes}
\label{sec6-2}

 Following \cite{Greenberg1991} we adopt three different regimes and their transitions, regime A: dominance by random motion; regime B: dominance by Keplerian shear motion; regime C: Keplerian shear dominance in a very thin disk.
 For the dispersion regime (dominance by random motion), where keplerian behavior is unimportant, we adopt the impact rate given by Eq.~(\ref{eq1-sec-2-2-1}). 

The transition between regime A and regime B is given by 
\begin{equation}
 \frac{(a_P+a_{T})}{2} \frac{(e_P+e_T)}{2}= 2.5 R_{H_T},
\label{eq2-sec-2-2-1}
\end{equation}
where $R_{H_T}$ is the Hill radius of the target, and $a_P,e_P$ and $a_T,e_T$ are the semi-major axis and eccentricities of the projectile and the target, respectively. Then, for larger values of $e$ and $a$ the system is dominated by random motion (regime A) and for smaller values of these orbital parameters the system is in the keplerian shear regime (regime B). If $a_p i_p < R_{G}$, being $R_{G}$ the target's gravitational diameter and $i_p$ the inclination of the projectile's orbit, the particles are in regime C, wherein still dominates the keplerian shear motion but the system is two dimensional. 

For regimes B and C the relative velocity is given by
\begin{equation}
v= 0.58 (2 \mu^{1/15}- 1.27)^{1/2} \Delta a, 
\label{eq3-sec-2-2-1}
\end{equation}
where $\mu = M_p/M_{\odot}$ and $\Delta a = 2.5 R_{H_T}$. The impact rate of regime B follows
\begin{eqnarray}
 \text{Impact rate}|_{\text{B}}= \pi R_P^2 \left( 1 + b \frac{V_{esc}^2}{v^2} \right)^{1/2} \dfrac{\sigma \left(2.5 R_{H_T}\right)^2 1.125 \omega}{\dfrac{a_P+a_T}{2} 4 a_{P} i_{P} \mu^{2/5} M_P},
\label{eq4-sec-2-2-1}
\end{eqnarray}
where $\sigma$ is the projectiles surface density and $\omega$ is the keplerian frequency using $a=(a_P+a_T)/2$.

Finally, the impact rate of regime C is given by 

\begin{equation}
\text{Impact rate}|_{\text{C}}= R_p  \left( 1 + b \frac{V_{esc}^2}{v^2} \right)^{1/2} \dfrac{\sigma (2.5 R_{H_T})^2 1.125 \omega}{\dfrac{(a_P+a_T)}{2} \mu^{2/5} M_p}.
\label{eq5-sec-2-2-1}
\end{equation}

\subsection{Catastrophic impact energy threshold}
\label{sec6-3}

 From SHP simulations, \cite{BenzAsphaug1999} found that $Q_{D}^*$ can be expressed by the functional form 
\begin{equation}
Q_{D}^* = Q_{0} \left(\frac{ R_T}{1~\text{cm}} \right)^a + B \rho  \left( \frac{R_T}{1~\text{cm}} \right)^b, 
\label{eq1-sec2-2-2}
\end{equation}
being $Q_{0}, B, a$, and $b$ parameters that depend on the properties of the material and on the impact velocity over the target, and $\rho$ the density of the non-porous planetesimals (in this work we adopted $\rho= 1.5~\text{gr}/\text{cm}^3$). They performed simulations for non-porous basalts at 3~km/s and 5~km/s, and for non-porous ices at 0.5~km/s and 3 km/s. Later, \cite{Benz2000}, performed new SPH simulations to calculate $Q_{D}^*$ for non-porous basalts at low impact velocities (between 20~m/s and 30 m/s) finding that targets impacted at such low impact velocities are weaker than targets impacted at greater velocities. Despite the fact that \cite{Benz2000} did not provide a functional form for $Q_{D}^*$, we fit the results of their simulations adopting the same functional form proposed by \cite{BenzAsphaug1999}. The Tab.~\ref{tab1-sec2-2-2} summarizes the values of the free parameters that determine the values of $Q_{D}^*$ used in this work for different types of materials and different impact velocities.

\begin{table}[t]
\caption{Free parameters that determine $Q_{D}^*$ for different types of materials and different impact velocities from \cite{BenzAsphaug1999} and \cite{Benz2000}.}
\label{tab1-sec2-2-2}    
\centering  
{\scriptsize  
\begin{tabular}{c c c c c }   
\hline\hline             
  & $Q_{0}$ & $B$ & $a$ & $b$  \\  
  & erg/g  & erg~cm$^3$/seg$^2$ & & \\
\hline                       
 Basalts at 5~km/s & 9.0e7 & 0.5 & -0.36 & 1.36 \\
\hline                       
Basalts at 3~km/s & 3.5e7 & 0.3 & -0.38 & 1.36 \\
\hline                       
Basalts at $\sim$~25~m/s & 1.22505e7 & 6.3e-8 & -0.305095 & 2.27386 \\
\hline
Ices at 3~km/s & 1.6e7 & 1.2 & -0.39 & 1.26 \\
\hline
Ices at 0.5~km/s & 7.0e7 & 2.1 & -0.45 & 1.19 \\
\hline\hline             
\end{tabular}
}
\end{table}

\end{appendix}

\end{document}